\documentclass[12pt]{article}
\usepackage{amsfonts}

\usepackage{graphicx}
\usepackage{amsmath}
\usepackage{float}

\numberwithin{equation}{section} \numberwithin{figure}{section}
\input{tcilatex}

\begin{document}

\title{Entropic N-bound and Maximal Mass Conjecture Violations in Four
Dimensional Taub-Bolt(NUT)-dS Spacetimes}
\author{R. Clarkson\thanks{
Email: rick@avatar.uwaterloo.ca}, A. M. Ghezelbash\thanks{
Email: amasoud@avatar.uwaterloo.ca}, R. B. Mann\thanks{
Email: mann@avatar.uwaterloo.ca} \\
Department of Physics, University of Waterloo, \\
Waterloo, Ontario N2L 3G1, CANADA}
\maketitle

\begin{abstract}
We show that the class of four-dimensional Taub-Bolt(NUT) spacetimes with
positive cosmological constant for some values of NUT charges are stable and
have entropies that are greater than that of de Sitter spacetime, in
violation of the entropic N-bound conjecture. We also show that the maximal
mass conjecture, which states \textit{"any asymptotically dS spacetime with
mass greater than dS has a cosmological singularity" }, can be violated as
well. Our calculation of conserved mass and entropy is based on an extension
of the path integral formulation to asymptotically de Sitter spacetimes.
\end{abstract}

\section{Introduction}

The set of conserved quantities associated with a given physical system is
one of its most fundamental features. If the system consists of a spacetime
that is either asymptotically flat (aF) or asymptotically anti de Sitter
(aAdS) these quantities are generally well understood. In the former case
the conserved quantities are the $d(d+1)/2$ conserved charges corresponding
to the Poincare generators in $d$-dimensions. \ In the latter case the
situation is a bit more problematic since the conserved charges have
supertranslation-like ambiguities (due to the coordinate dependence of the
formalism). In either case such quantities have been defined relative to an
auxiliary spacetime, in which the boundary of the spacetime of interest must
be embedded in a reference spacetime. In recent years progress in this area
has been made by incorporating expectations from the AdS/CFT correspondence.
\ By introducing additional surface terms that are functionals of geometric
invariants on the boundary \cite{balakraus}, an alternative approach was
developed for computing conserved quantities associated with aAdS spacetimes
that was free of the aforementioned difficulties.

Both aF and aAdS spacetimes have spatial infinity, a property that plays an
important role in the construction of conserved charges for both of these
cases. \ However, asymptotically de Sitter (adS) spacetimes do not have a
spatial infinity, and so present an interesting puzzle in the calculation of
conserved quantities. Such spacetimes also have no global timelike Killing
vector; rather the norm of the Killing vector changes sign depending on
which side of the horizon one is considering. Calculations for conserved
charges and actions/entropies have been carried out for pure and
asymptotically de Sitter spacetimes inside the cosmological horizon where
the Killing vector is timelike \cite{GibbonsHawking1}. \ Outside of this
horizon the physical meaning of energy and other conserved quantities is not
clear; for example to construct the energy one could use the conformal
Killing vector \cite{Shiro}.\bigskip

Recently a novel prescription was proposed for computing conserved charges
(and associated boundary stress tensors) of adS spacetimes from data at
early or late time infinity \cite{bala}. The method is analogous to the
Brown-York prescription in asymptotically AdS spacetimes \cite%
{balakraus,brown,BCM,ivan}, suggesting a holographic duality similar to the
AdS/CFT correspondence. The specific prescription in ref. \cite{bala} (which
has been employed previously by others but in more restricted contexts \cite%
{Klem, Myung}) presented the counterterms on spatial boundaries at early and
late times that yield a finite action for asymptotically dS spacetimes in $%
3,4,5$ dimensions. By carrying out a procedure analogous to that in the AdS
case \cite{balakraus,BCM}, one could compute the boundary stress tensor on
the spacetime boundary, and consequently a conserved charge interpreted as
the mass of the asymptotically dS spacetime could be calculated. \bigskip

The conserved charge associated with the Killing vector $\partial /\partial
t $ -- now spacelike outside of the cosmological horizon -- was interpreted
as the conserved mass \cite{bala}. \ Employing this definition, the authors
of ref. \cite{bala} were led to the conjecture that \textit{any
asymptotically dS spacetime with mass greater than dS has a cosmological
singularity. }\ We shall refer to this conjecture as the maximal mass
conjecture. As stated, the conjecture is in need of clarification before a
proof could be considered, but roughly speaking it means that the conserved
mass of any physically reasonable adS spacetime must be negative (i.e. less
than the zero value of pure dS spacetime). \ This has been verified for
topological dS solutions and their dilatonic variants \cite{cai} and for the
Schwarzschild-de Sitter (SdS) black hole up to dimension nine \cite{GM}. The
maximal mass conjecture was based in part on the Bousso N-bound \cite{bousso}%
, another conjecture stating that \textit{any asymptotically dS spacetime
will have an entropy no greater than the entropy }$\pi \ell ^{2}$\textit{\
of pure dS with cosmological constant }$\Lambda =3/\ell ^{2}$\textit{\ in }$%
(3+1)$\textit{\ dimensions}.

\bigskip

We have found, however, that locally asymptotically de Sitter spacetimes --
with NUT charge $n$ -- provide counterexamples to both of these conjectures
under certain circumstances \cite{dsnutshort}.{\Large \ }In this paper we
explore this situation in detail, illustrating the circumstances under which
the conjectures are and are not satisfied. \ We demonstrate that locally
asymptotically de Sitter spacetimes with NUT charge in $(3+1)$\ dimensions
can violate both the N-bound and the maximal mass conjecture. \ While it has
been shown recently that the class of stable spacetimes in the form of $
dS_{p}\times S^{q}$\ in more than four dimensions ($p+q>4$), have entropy
greater than that of de Sitter spacetime (in violation of the N-bound) \cite%
{boussomyers}, to our knowledge this is the first demonstration of
spacetimes that violate the maximal mass conjecture.

The paper proceeds as follows. In section \ref{sec:general}, we will outline
and review the procedure for calculating the conserved mass and entropy, and
derive the general expressions for these quantities in ($d+1$) dimensions.
Since there are several different ways of writing the metric, depending on
which set of Wick rotations is chosen, these calculations will in the next
section be shown for all such choices. In section \ref{sec:4d}, we will
analyze the quantities in ($3+1$) dimensions, and compare our various
approaches - also, we will demonstrate where these solutions violate the
Bousso bound and the maximal mass conjectures, as both of these have been
formulated for ($3+1$) dimensions. In section \ref{sec:6d}, we will analyze
the conserved mass and entropy in ($5+1$) dimensions. NUT-charged spacetimes
of dimensionality $4k$\ are qualitatively similar to the ($3+1$) dimensional
case whereas those of dimensionality $4k+2$\ are qualitatively similar to
the ($5+1$) dimensional case.

\section{Path-Integration and Asymptotically de Sitter Spacetimes}

\label{sec:general}

\bigskip We begin with the path-integral approach, for which 
\begin{equation}
\left\langle g_{2},\Phi _{2},S_{2}|g_{1},\Phi _{1},S_{1}\right\rangle =\int
D \left[ g,\Phi \right] \exp \left( iI\left[ g,\Phi \right] \right)
\label{PI1}
\end{equation}%
represents the amplitude to go from a state with metric and matter fields $%
\left[ g_{1},\Phi _{1}\right] $ on a surface $S_{1}$ to a state with metric
and matter fields $\left[ g_{2},\Phi _{2}\right] $ on a surface $S_{2}$. \
The quantity $D\left[ g,\Phi \right] $ is a measure on the space of all
field configurations and $I\left[ g,\Phi \right] $ is the action taken over
all fields having the given values on the surfaces $S_{1}$ and $S_{2}$. \
For asymptotically flat and asymptotically anti de Sitter spacetimes these
surfaces are joined by timelike tubes of some finite mean radii, so that the
boundary and the region contained within are compact. \ In the limit that
the larger mean radius becomes infinite and the smaller mean radius vanishes
one obtains the amplitude for the entire spacetime and matter fields to
evolve from $\left[ g_{1},\Phi _{1},S_{1}\right] $ to $\left[ g_{2},\Phi
_{2},S_{2}\right] $.

In the case of asymptotically $(d+1)$-dimensional de Sitter spacetimes the
situation is somewhat different. \ We replace the surfaces $S_{1},S_{2}$
with histories $H_{1},H_{2}$\ that have spacelike unit normals and are
surfaces that form the timelike boundaries of a given spatial region; they
therefore describe particular histories of $d$-dimensional subspaces of the
full spacetime. \ The amplitude (\ref{PI1}) becomes 
\begin{equation}
\left\langle g_{2},\Phi _{2},H_{2}|g_{1},\Phi _{1},H_{1}\right\rangle =\int
D \left[ g,\Phi \right] \exp \left( iI\left[ g,\Phi \right] \right)
\label{PI2}
\end{equation}%
and describes quantum correlations between differing histories $\left[
g_{1},\Phi _{1}\right] $ and $\left[ g_{2},\Phi _{2}\right] $\ of metrics
and matter fields. The correlation between a history $\left[ g_{2},\Phi
_{2},H_{2}\right] $\ with a history $\left[ g_{1},\Phi _{1},H_{1}\right] $
is obtained from the modulus squared of this amplitude. \ The surfaces $%
H_{1},H_{2}$\ are joined by spacelike tubes at some initial and final times,
so that the boundary and interior region are compact. In the limit that
these times approach past and future infinity one obtains the correlation
between the complete histories $\left[ g_{1},\Phi _{1},H_{1}\right] $ and $%
\left[ g_{2},\Phi _{2},H_{2}\right] $. \ This correlation is given by
summing over all metric and matter field configurations that interpolate
between these two histories. The quantity $\left\langle g_{2},\Phi
_{2},S_{2}|g_{1},\Phi _{1},S_{1}\right\rangle $\ depends only on the metrics 
$g_{1},g_{2}$\ and the fields $\Phi _{1},\Phi _{2}$\ on the initial and
final surfaces $S_{1},S_{2}$\ and not on a special metric $g_{i}$\ and
matter fields $\Phi _{i}$\ on an intermediate surface $S_{i}$\ embedded in
spacetime; in the evaluation of the amplitude $\left\langle g_{2},\Phi
_{2},S_{2}|g_{1},\Phi _{1},S_{1}\right\rangle $ all the different possible
configurations $g_{i},\Phi _{i}$ in spacetime are summed$.$\ Similarly, the
quantity $\left\langle g_{2},\Phi _{2},H_{2}|g_{1},\Phi
_{1},H_{1}\right\rangle $ depends only on the hypersurfaces $H_{1}$\ and $%
H_{2}$\ and the metrics and matter fields over these hypersurfaces and not
on any special hypersurface in the spacetime, between the hypersurfaces $%
H_{1}$\ and $H_{2}.$\ 

\bigskip

The action can be decomposed into three distinct parts 
\begin{equation}
I=I_{B}+I_{\partial B}+I_{ct}  \label{actiongeneral}
\end{equation}%
where the bulk ($I_{B}$) and boundary ($I_{\partial B}$) terms are the usual
ones, given by 
\begin{eqnarray}
I_{B} &=&\frac{1}{16\pi }\int_{\mathcal{M}}d^{d+1}x~\sqrt{-g}\left(
R-2\Lambda +\mathcal{L}_{M}(\Phi )\right)  \label{actionbulk} \\
I_{\partial B} &=&-\frac{1}{8\pi }\int_{\partial \mathcal{M}^{\pm }}d^{d}x~ 
\sqrt{\gamma ^{\pm }}\Theta ^{\pm }  \label{actionboundary}
\end{eqnarray}%
where $\partial \mathcal{M}^{\pm }$ represents future/past infinity, and $
\int_{\partial \mathcal{M}^{\pm }}=\int_{\partial \mathcal{M}^{-}}^{\partial 
\mathcal{M}^{+}}$ represents an integral over a future boundary minus a past
boundary, with the respective metrics $\gamma ^{\pm }$ and extrinsic
curvatures $\Theta ^{\pm }$ (working in units where $G=1$). The quantity $ 
\mathcal{L}_{M}(\Phi )$ in (\ref{actionbulk}) is the Lagrangian for the
matter fields, which we won't be considering here. The bulk action is over
the $\left( d+1\right) $ dimensional manifold $\mathcal{M}$, and the
boundary action is the surface term necessary to ensure well-defined
Euler-Lagrange equations.

For an asymptotically dS spacetime, the boundary $\partial \mathcal{M}$ will
be a union of Euclidean spatial boundaries at early and late times. The
necessity of the boundary term (\ref{actionboundary}) can also be understood
from the path-integral viewpoint by considering the joint correlation
between histories $\left[ g_{1},\Phi _{1},H_{1}\right] $ and $\left[
g_{2},\Phi _{2},H_{2}\right] $ and also between $\left[ g_{2},\Phi
_{2},H_{2} \right] $ and $\left[ g_{3},\Phi _{3},H_{3}\right] $. The
correlation between $\left[ g_{1},\Phi _{1},H_{1}\right] $ and $\left[
g_{3},\Phi _{3},H_{3}\right] $ should be given by summing over the products
of correlations with all possible intermediate histories $\left[ g_{2},\Phi
_{2},H_{2}\right] $: 
\begin{equation}
\left\langle g_{3},\Phi _{3},H_{3}|g_{1},\Phi _{1},H_{1}\right\rangle
=\sum_{2}\left\langle g_{3},\Phi _{3},H_{3}|g_{2},\Phi
_{2},H_{2}\right\rangle \left\langle g_{2},\Phi _{2},H_{2}|g_{1},\Phi
_{1},H_{1}\right\rangle  \label{PI3}
\end{equation}
which will hold provided 
\begin{equation}
I\left[ g_{13},\Phi \right] =I\left[ g_{12},\Phi \right] +I\left[
g_{23},\Phi \right]  \label{action123}
\end{equation}
where $g_{12}$ is the metric of the spacetime region between histories $%
H_{1} $ and $H_{2}$ and $g_{23}$ is the metric of that between histories $%
H_{2}$ and $H_{3}$. The metric $g_{13}$ is the metric of the full spacetime
between histories $H_{1}$ and $H_{3}$ obtained by joining the two regions.
In general the metrics \ $g_{12}$ and $g_{23}$ will have different spacelike
normal derivatives, yielding delta-function contributions to the Ricci
tensor proportional to the difference between the extrinsic curvatures of
the history $H_{2}$ in the metrics $g_{12}$ and $g_{23}$. \ The boundary
term $I_{\partial B}$ compensates for these discontinuities, and so ensures
that the relation (\ref{action123}) holds.

We next turn to a consideration of the counter-term action $I_{ct}$ in (\ref%
{actiongeneral}). In the context of the dS/CFT correspondence conjecture, it
appears due to the counterterm contributions from the boundary quantum CFT %
\cite{balakraus,CFTref}. The existence of such terms in de Sitter space is
suggested by analogy with the AdS/CFT correspondence, which posits the
relationship 
\begin{equation}
Z_{\text{AdS}}[\gamma ,\Psi _{0}]=\int_{[\gamma ,\Psi _{0}]}D\left[ g\right]
D\left[ \Psi \right] e^{-I\left[ g,\Psi \right] }=\left\langle \exp \left(
\int_{\partial \mathcal{M}{_{d}}}d^{d}x\sqrt{g}\mathcal{O}_{[\gamma ,\Psi
_{0}]}\right) \right\rangle =Z_{CFT}[\gamma ,\Psi _{0}]  \label{PAR}
\end{equation}%
between the partition function of the field theory on AdS$_{d+1}$ and its
quantum conformal field theory on its boundary. The counter-term action $%
I_{ct}$ in eq. (\ref{actiongeneral}) appears for similar reasons: \ the
quantum CFT at future/past infinity is expected to have counterterms whose
values can only depend on geometric invariants of these spacelike surfaces.
The counterterm action can be shown to be universal for both the AdS and dS
cases by re-writing the Einstein equations in Gauss-Codacci form, and then
solving them in terms of the extrinsic curvature and its derivatives to
obtain the divergent terms; these will cancel the divergences in the bulk
and boundary actions \cite{GM,KrausLarsenSieb}. It can be generated by an
algorithmic procedure, without reference to a background metric, and yields
finite values for conserved quantities that are intrinsic to the spacetime.
The result of employing this procedure in de Sitter spacetime is \cite{GM} 
\begin{eqnarray}
I_{\text{ct}} &=&-\int d^{d}x\sqrt{\gamma }\left\{ -\frac{d-1}{\ell }+\frac{%
\ell \mathsf{\Theta }\left( d-3\right) }{2(d-2)}\mathsf{R}-\frac{\ell ^{3}%
\mathsf{\Theta }\left( d-5\right) }{2(d-2)^{2}(d-4)}\left( \mathsf{R}_{ab}%
\mathsf{R}^{ab}-\frac{d}{4(d-1)}\mathsf{R}^{2}\right) \right.  \notag \\
&&-\frac{\ell ^{5}\mathsf{\Theta }\left( d-7\right) }{(d-2)^{3}(d-4)(d-6)}%
\left( \frac{3d+2}{4(d-1)}\mathsf{RR}^{ab}\mathsf{R}_{ab}-\frac{d(d+2)}{%
16(d-1)^{2}}\mathsf{R}^{3}\right.  \notag \\
&&\left. -2\mathsf{R}^{ab}\mathsf{R}^{cd}\mathsf{R}_{acbd}\left. -\frac{d}{%
4(d-1)}\nabla _{a}\mathsf{R}\nabla ^{a}\mathsf{R}+\nabla ^{c}\mathsf{R}%
^{ab}\nabla _{c}\mathsf{R}_{ab}\right) +\ldots \right\}  \label{actionct}
\end{eqnarray}%
with $\mathsf{R}$ the curvature of the induced metric $\gamma $ and $\Lambda
={\textstyle\frac{d(d-1)}{2\ell ^{2}}}$. The step-function $\mathsf{\Theta }%
\left( x\right) $ is unity provided $x>0$ and vanishes otherwise, thereby
ensuring that the series only contains the terms necessary to cancel
divergences and no more. Hence, for example, in four ($d=3$) dimensions,
only the first two terms appear, and only these are needed to cancel
divergent behavior in $I_{B}+I_{\partial B}$ near past and future infinity.

If the boundary geometries have an isometry generated by a Killing vector $%
\xi ^{\pm \mu }$, then it is straightforward to show that $T_{ab}^{\pm }\xi
^{\pm b}$ is divergenceless, from which it follows that

\begin{equation}
\mathfrak{Q}{}^{\pm }=\oint_{\Sigma ^{\pm }}d^{d-1}\varphi ^{\pm }\sqrt{%
\sigma ^{\pm }}n^{\pm a}T_{ab}^{\pm }\xi ^{\pm b}  \label{Qcons}
\end{equation}%
is conserved between histories of constant $t$, whose unit normal is given
by $n^{\pm a}$. The $\varphi ^{a}$ are coordinates describing closed
surfaces $\Sigma $, where we write the boundary metric(s) of the spacelike
tube(s) as

\begin{equation}
h_{ab}^{\pm }d\hat{x}^{\pm a}d\hat{x}^{\pm b}=d\hat{s}^{\pm 2}=N_{T}^{\pm
2}dT^{2}+\sigma _{ab}^{\pm }\left( d\varphi ^{\pm a}+N^{\pm a}dT\right)
\left( d\varphi ^{\pm b}+N^{\pm b}dT\right)  \label{hmetric}
\end{equation}%
where $\nabla _{\mu }T$ is a spacelike vector field that is the analytic
continuation of a timelike vector field. Physically this means that a
collection of observers on the hypersurface all observe the same value of $%
{}Q$ provided this surface has an isometry generated by $\xi ^{b}$.
Alternatively it means that for any two histories $\left[ g_{1},\Phi
_{1},H_{1}\right] $\ and $\left[ g_{2},\Phi _{2},H_{2}\right] $, the value
of $Q$\ is the same for each provided $\xi $\ is a Killing vector on the
surface $\Sigma $. Note that unlike the asymptotically flat and AdS cases
the surface $\Sigma $\ does not enclose anything; rather it is the boundary
of the class of histories that interpolate between $H_{1}$\ and $H_{2}$. \
In this sense $Q$\ is associated only with this boundary and not with the
class of histories that it bounds. This is analogous to the situation in
asymptotically flat and AdS spacetimes, in which conserved quantities can be
associated with surfaces whose interiors have no isometries \cite{ivan}.

If $\partial /\partial t$ is itself a Killing vector, then we define 
\begin{equation}
\mathfrak{M}{}^{\pm }=\oint_{\Sigma ^{\pm }}d^{d-1}\varphi ^{\pm }\sqrt{%
\sigma ^{\pm }}N_{T}^{\pm }n^{\pm a}n^{\pm b}T_{ab}^{\pm }  \label{Mcons}
\end{equation}%
as the conserved mass associated with the future/past surface $\Sigma ^{\pm
} $\ at any given point $t$ on the spacelike future/past boundary. This
quantity changes with the cosmological time $\tau $. Since all
asymptotically de Sitter spacetimes must have an asymptotic isometry
generated by $\partial /\partial t$, there is at least the notion of a
conserved total mass ${}\mathfrak{M}^{\pm }$ for the spacetime in the limit
that $\Sigma ^{\pm }$ are future/past infinity. Similarly the quantity 
\begin{equation}
\mathfrak{J}^{\pm a}=\oint_{\Sigma ^{\pm }}d^{d-1}\varphi ^{\pm }\sqrt{%
\sigma ^{\pm }}\sigma ^{\pm }{}^{ab}n^{\pm c}T_{bc}^{\pm }  \label{Jcons}
\end{equation}%
can be regarded as a conserved angular momentum associated with the surface $%
\Sigma ^{\pm }$\ if the surface has an isometry generated by $\partial
/\partial \phi ^{\pm a}$.

We turn now to an approach for evaluating and interpreting the path
integral. The action (\ref{actiongeneral}) will be real for Lorentzian
metrics and real matter fields, and so the path integral will be oscillatory
and so will not converge. \ To set the stage for the calculations we perform
we briefly review the path-integral approach to quantum gravity and its
relationship to gravitational thermodynamics for asymptotically flat or
asymptotically AdS spacetimes. \ 

Consider a scalar quantum field $\phi $ -- the amplitude for going from a
state $|{t_{1},\phi _{1}}\rangle $ to $|{t_{2},\phi _{2}}\rangle $ can be
expressed as an integral 
\begin{equation}
\langle {t_{2},\phi _{2}}|{t_{1},\phi _{1}}\rangle =\int_{1}^{2}d[\phi ]e^{%
\text{i}I[\phi ]}  \label{qmamplitude1}
\end{equation}%
over all possible intermediate field configurations between the initial and
final states. However, this amplitude can also be expressed as 
\begin{equation}
\langle {t_{2},\phi _{2}}|{t_{1},\phi _{1}}\rangle =\langle {\phi _{2}}|e^{-%
\text{i}H(t_{2}-t_{1})}|{\phi _{1}}\rangle  \label{qmamplitude2}
\end{equation}%
where $H$ is the Hamiltonian. By imposing the periodicity condition $\phi
_{1}=\phi _{2}$ for $t_{2}-t_{1}=-i\beta $, we sum over $\phi _{1}$to obtain 
\begin{equation}
\text{Tr}(\exp (-\beta H))=\int d[\phi ]e^{-\hat{I}[\phi ]}
\label{partition1}
\end{equation}%
The right-hand side is now a Euclidean path integral over all field
configurations intermediate between the periodic boundary conditions because
of the Wick rotation of the time coordinate, where $\hat{I}$ is the
Euclidean action. Inclusion of gravitational effects can be carried out as
described above, by considering the initial state to include a metric on a
surface $S_{1}$ at time $t_{1}$ evolving to another metric on a surface $%
S_{2}$ at time $t_{2}$, yielding the relation (\ref{PI1}).

\bigskip

Note that the left-hand side of (\ref{partition1}) is simply the partition
function $Z$ for the canonical ensemble for a field at temperature $\beta
^{-1}$. This connection with standard thermodynamic arguments \cite{Pathria}
can be seen as follows. We start with the canonical distribution 
\begin{equation}
P_{r}\equiv \frac{<n_{r}>}{\mathcal{N}}=\frac{e^{-\beta E_{r}}}{%
\sum_{r}e^{-\beta E_{r}}}  \label{Pr}
\end{equation}%
with $\beta $ determined by considering the average total energy $M$%
\begin{equation}
M=\frac{\sum_{r}E_{r}e^{-\beta E_{r}}}{\sum_{r}e^{-\beta E_{r}}}=-\frac{%
\partial }{\partial \beta }\ln \left\{ \sum_{r}e^{-\beta E_{r}}\right\} =-%
\frac{\partial }{\partial \beta }\ln Z  \label{Mnormal}
\end{equation}%
Also, the Helmholtz free energy $W=M-TS$ can be rearranged so that 
\begin{subequations}
\begin{eqnarray}
M=W+TS &=&W-T\left( \frac{\partial W}{\partial T}\right) _{N,V}=-T^{2}\left[ 
\frac{\partial }{\partial T}\left( \frac{W}{T}\right) \right] _{N,V}
\label{Mnormal2} \\
&=&\frac{\partial }{\partial \beta }\left( \beta W\right)  \label{Mnormal3}
\end{eqnarray}%
Comparing (\ref{Mnormal}) and (\ref{Mnormal3}), we get 
\end{subequations}
\begin{equation}
-\beta W=\ln \left\{ \sum_{r}e^{-\beta E_{r}}\right\} =\ln Z
\label{GDnormal}
\end{equation}%
which can be interpreted as describing the partition function of a
gravitational system at temperature $\beta ^{-1}$ contained in a (spherical)
box of finite radius.

\bigskip

We therefore compute $Z$ using an analytic continuation of the action in (%
\ref{PI1}) so that the axis normal to the surfaces $S_{1},S_{2}$ is rotated
clockwise by $\frac{\pi }{2}$ radians into the complex plane \cite%
{GibbonsHawking1} (i.e. by rotating the time axis so that $t\rightarrow iT$
) in order to obtain a Euclidean signature. \ The positivity of the
Euclidean action ensures a convergent path integral in which one can carry
out any calculations (of action, entropy, etc.). The presumed physical
interpretation of the results is then obtained by rotation back to a
Lorentzian signature at the end of the calculation. \ 

\bigskip

In the asymptotically de Sitter case these arguments require a greater
degree of care because the action is in general negative definite near past
and future infinity (outside of a cosmological horizon). The natural
strategy would appear to be to analytically continue the coordinate
orthogonal to the histories $\left[ g_{1},\Phi _{1},H_{1}\right] $ and $%
\left[ g_{2},\Phi _{2},H_{2}\right] $ to complex values by rotating the axis
normal to the histories $H_{1},H_{2}$\ anticlockwise by $\frac{\pi }{2}$
radians into the complex plane. The action becomes pure imaginary and so $%
\exp \left( iI\left[ g,\Phi \right] \right) \longrightarrow \exp \left( +%
\hat{I}\left[ g,\Phi \right] \right) $, yielding a convergent path integral 
\begin{equation}
Z^{\prime }=\int e^{+\hat{I}}  \label{Partitionaction}
\end{equation}%
since $\hat{I}<0$. \ Furthermore, since we want a converging partition
function, we must change (\ref{Mnormal}) to 
\begin{equation}
M=+\frac{\partial }{\partial \beta }\ln \left\{ \sum_{r}e^{+\beta
E_{r}}\right\} =+\frac{\partial }{\partial \beta }\ln Z^{\prime }
\label{Mout}
\end{equation}%
Now comparing (\ref{Mout}) with (\ref{Mnormal3}) (since (\ref{Mnormal2},\ref%
{Mnormal3}) won't change) we will obtain%
\begin{equation}
+\beta W=\ln \left\{ e^{+\beta E_{r}}\right\} =\ln Z^{\prime }
\label{GDout1}
\end{equation}%
In the semi-classical approximation this will lead to $\ln Z^{\prime
}=+I_{cl}$. Substituting this and (\ref{Mnormal2}) into (\ref{GDout1}), 
\begin{eqnarray}
\beta \left( M-TS\right) &=&+I_{cl}  \notag \\
\beta M-S &=&I_{cl}  \notag \\
S &=&\beta M-I_{cl}  \label{GDoutfinal}
\end{eqnarray}

As before, the presumed physical interpretation of the results is then
obtained by rotation back to a Lorentzian signature at the end of the
calculation. However there is an ambiguity here that is not present in the
asymptotically flat and AdS cases. This occurs because outside the horizon,
near past and future infinity, the signature of any asymptotically dS
spacetime becomes $(+,-,+,+)$, and so the spacelike boundary tubes naturally
have Euclidean signature. This leads to two possible approaches in
evaluating physical quantities.

In the first approach one proceeds in a manner similar to the asymptotically
flat and AdS cases, performing all calculations after anticlockwise rotation
into the complex plane of the spacelike axis normal to histories. This
involves not only a complex rotation of the (spacelike) $t$
coordinate\thinspace\ ($t\rightarrow iT$), but also an analytic continuation
of any rotation and NUT charge parameters, yielding a metric of signature ($%
-,-,+,+,\ldots $). In the calculation of the action, this will give rise to
a negative action, and hence a negative definite energy. Our argument for
this approach is that it is not the Euclidean signature of the metric that
is important, but rather the convergence of the path integral and of the
partition function. We also periodically identify $T$ with period $\beta $
(given by the surface gravity of the cosmological horizon of the $\left(
-,-\right) $ section)\ to ensure the absence of conical singularities. We
shall refer to this approach as the C-approach, since it involves a rotation
into the complex plane.

In the second approach, we note that at future infinity $\partial /\partial
t=\partial _{t}$ is asymptotically a spacelike Killing vector. This suggests %
\cite{GM} that rotation into the complex plane is merely formal device whose
function is to establish the relationship (\ref{GDoutfinal}); it is not
necessary for computational purposes. One can simply evaluate the action at
future infinity, imposing periodicity in $t$, consistent with regularity at
the cosmological horizon (given by the surface gravity of the cosmological
horizon of the $\left( +,-\right) $ section). There is no need to
analytically continue either rotation parameters or NUT to complex values.
We shall refer to this approach as the R-approach, since no quantities are
analytically continued into the complex plane. \ 

\bigskip

In adS spacetimes with NUT charge there is an additional periodicity
constraint in $t$ that arises from demanding that no Misner-string
singularities appear in the spacetime. \ When incorporated with the
periodicity $\beta $, this yields an additional consistency criterion that
relates the mass and NUT parameters, the two solutions of which produce
generalizations of asymptotically flat Taub-Bolt space to the asymptotically
de Sitter case \cite{dsnutshort}. These solutions can be classified by the
dimensionality of the fixed point sets of the Killing vector $\xi =\partial
/\partial t$\ that generates a $U(1)$ isometry group. \ If this fixed point
set dimension is $\left( d-1\right) $ the solution is called a Bolt
solution; if the dimensionality is less than this then the solution is
called a NUT solution. We shall see that in the C-approach this yields a
dS-NUT solution analogous to the AdS-NUT case, as well as the Bolt
solutions, whereas the R-approach yields Bolt solutions only. Both of these
versions of the Taub-NUT-dS spacetime are solutions to the Einstein
equations.

\bigskip

Our proposal (\ref{PI2}) for extending the path-integral formalism for
quantum gravity to describe quantum correlations between differing histories
(as opposed to quantum amplitudes between differing spacelike slices) can be
physically motivated in the following way. \ Consider for definiteness an
adS spacetime with cylindrical topology. A given history can be interpreted
as the collection of worldlines of a set of observers on a compact slice at
a given point $t$\ performing a variety of experiments in a (cosmologically)
evolving universe. \ The choice of initial point of their history is
determined by the time at which they began their experiments and the final
point corresponds to the time at which they completed their experiments. \
Their entire history determines a causal diamond given by the intersection
of the causal future of their initial point with the causal past of their
final point. \ If their experiments begin at past infinity and end at future
infinity they they have obtained the maximal amount of information that they
can empirically access.

The path integral (\ref{PI2}) then describes the quantum correlation between
the information these observers collected within their causal diamond with
that obtained by another class of observers at some other point $t^{\prime }$%
. \ A given class of observers could split up, choosing the same initial
point but have differing intermediate histories. Similarly, two classes of
observers could choose to meet at some final point, or a given class could
split up at some initial time and reunite at some final time. \ In all such
cases the interpretation of the path integral (\ref{PI2}) would be that it
describes the quantum correlations between the information contained in
their observations. The modulus squared of the amplitude would represent the
probability that the information collected from one history is correlated
with that of another.

\bigskip

We note also that although we have derived eq. (\ref{GDoutfinal}) from the
path integral formalism, its thermodynamic interpretation remains to be
fully understood. However it would seem reasonable to expect that
gravitational entropy $S$ is generated by the presence of past/future
cosmological horizons, and that the entropy in eq. (\ref{GDoutfinal}) counts
the degrees of freedom hidden behind such horizons. \ Of course there is a
distinction between the entropy at past infinity and the entropy at future
infinity. \ A future cosmological horizon shields information from observers
at or near past infinity somewhat analogously to the manner in which a black
hole shields external observers from the information inside. \ They have the
option of actively probing for information behind the horizon, but only at
later times. Observers at or near future infinity cannot probe for
information from behind the past cosmological horizon; rather they can only
passively access it.

As an application of the relation (\ref{GDoutfinal}), consider the $(d+1)$%
-dimensional SdS spacetimes, with metric%
\begin{equation}
ds^{2}=-\frac{d\tau ^{2}}{\frac{\tau ^{2}}{\ell ^{2}}+\frac{2m}{\tau ^{d-2}}%
-1}+\left( \frac{\tau ^{2}}{\ell ^{2}}+\frac{2m}{\tau ^{d-2}}-1\right)
dt^{2}+\tau ^{2}d\Omega _{d-1}^{2}  \label{sdsmetric}
\end{equation}%
The mass is given by $M_{^{d+1}}=-\frac{V_{d-1}}{16\pi }\{2(d-1)m-C_{d}\}$%
\textbf{\ }where $V_{d}$\ is the volume of the unit $d$-sphere, $C_{d}$ is
the Casimir energy which is non-vanishing for even $d$\textbf{\ }and we
obtain \cite{GM}%
\begin{equation}
S_{d+1}=\frac{\beta \left( \tau _{+}^{d}-(d-2)m\ell ^{2}\right) V_{d-1}}{%
8\pi \ell ^{2}}=\frac{A_{d-1}}{4}  \label{sdsentropy}
\end{equation}%
for the entropy where $\tau _{+}$ is the largest root of the lapse function
and $m$\ is the mass parameter. The gravitational entropy $S_{d+1}$\ is a
positive monotonically increasing (decreasing) function of$\ \ $the
conserved total mass $M$ (mass parameter $m$) and so the N-bound is
satisfied. In the special case of $(2+1)$-dimensions, $S_{3}$\ is exactly
the same as Cardy formula \cite{bala,GM}. In higher dimensions, the
Gibbs-Duhem entropy (\ref{sdsentropy}) is less than the entropy associated
with the cosmological horizon, in agreement with the N-bound.

\section{General Considerations of NUT-charged Spacetimes}

The general form for the NUT-charged dS spacetime in ($d+1$) dimensions is
given by 
\begin{equation}
ds^{2}=V(\tau )\left( dt+nA\right) ^{2}-\frac{d\tau ^{2}}{V(\tau )}+(\tau
^{2}+n^{2})d\Gamma ^{2}  \label{TNDSgeneral}
\end{equation}%
where $~d=2k+1$ and $V(\tau )$ is given by the general formula 
\begin{equation}
V(\tau )=\frac{2m\tau }{(\tau ^{2}+n^{2})^{k}}-\frac{\tau }{(\tau
^{2}+n^{2})^{k}}\int_{\tau }ds\left[ \frac{(s^{2}+n^{2})^{k}}{s^{2}}-\frac{%
(2k+1)}{\ell ^{2}}\frac{(s^{2}+n^{2})^{k+1}}{s^{2}}\right]  \label{FtBoltgen}
\end{equation}%
with $n$ the non-vanishing NUT charge and $\Lambda =\frac{d(d-1)}{2\ell ^{2}}
$.

The one-form $A$ is a function of the coordinates $(\vartheta _{1},\phi
_{1},\cdot \cdot \cdot ,\vartheta _{k},\phi _{k})$ of the non-vanishing
compact base space of positive curvature (with metric $d\Gamma ^{2}$). The
coordinate $t$ parameterizes a circle $S^{1}$ Hopf-fibered over this space;
it must have periodicity $\frac{2(d+1)\pi \left| n\right| }{q}$ to avoid
conical singularities, where $q$ is a positive integer. The geometry of a
constant-$\tau $ surface is that of a Hopf fibration of $S^{1}$ over the
base space, which is a well defined spacelike hypersurface in spacetime
where $V(\tau )>0$ outside of the past/future cosmological horizons. The
spacelike Killing vector $\partial /\partial t$\ has a fixed point set where 
$V(\tau _{c})=0$\ whose topology is the same as that of the base space.

The general form of the base space is a combination of products of $S^{2}$
and $CP^{2}$, i.e. $\otimes _{i=1}^{s}S^{2}\otimes _{j=1}^{c}CP^{2}$ such
that $s+2c=k.$ The metric of $CP^{2}$ has the general form 
\begin{equation}
d\Sigma ^{2}=\frac{1}{(1+\frac{u^{2}}{6})^{2}}\{du^{2}+\frac{u^{2}}{4}(d\psi
+\cos \theta d\phi )^{2}\}+\frac{u^{2}}{4(1+\frac{u^{2}}{6})}(d\theta
^{2}+\sin ^{2}\theta d\phi ^{2})  \label{CP2}
\end{equation}%
for which the one-form $A$ is 
\begin{equation}
A=\frac{u^{2}}{2(1+\frac{u^{2}}{6})}(d\psi +\cos \theta d\phi )  \label{A}
\end{equation}%
whereas 
\begin{equation}
A=2\cos \theta d\phi  \label{Asph}
\end{equation}%
if the base space is a 2-sphere with metric $d\Omega ^{2}=d\theta ^{2}+\sin
^{2}\theta d\phi ^{2}$. \ For the general form $\otimes
_{i=1}^{s}S^{2}\otimes _{j=1}^{c}CP^{2}$ the one-form $A$ is a linear
combination of metrics of the forms (\ref{A}) and (\ref{Asph}).

\begin{figure}[tbp]
\centering        
\begin{minipage}[c]{.55\textwidth}
         \centering
         \includegraphics[width=\textwidth]{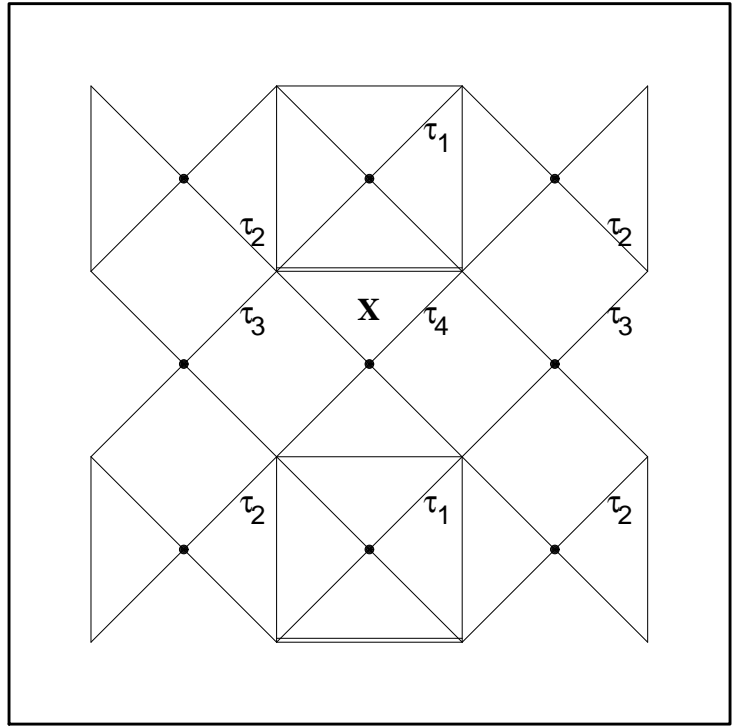}
         \label{PENROSE}
     \end{minipage}
\caption{The Penrose diagram of the TBdS spacetime. We denote the roots of $%
V $ by the increasing sequence $\protect\tau _{1}<0<\protect\tau _{2}<%
\protect\tau _{3}<\protect\tau _{4}=\protect\tau _{c}.$ The vertical and
horizontal lines are the $\protect\tau =0$ and the $\protect\tau =-\infty $
slices of the spacetime, respectively. The double line denotes $\protect\tau %
=+\infty $ and the solid black dots denote the quasiregular singularities of
the spacetime. Our calculation is performed outside the cosmological
horizon, located within the triangle denoted by ``X''. }
\end{figure}

The causal structure of TNdS spacetime can be understood by looking at a
typical Penrose diagram (Figure \ref{PENROSE}). For simplicity, we consider
a four-dimensional TNdS with an $S^{2}$ base space. We denote the roots of $%
V(\tau )$\ by the increasing sequence $\tau _{1}<0<\tau _{2}<\tau _{3}<\tau
_{4}=\tau _{c}.$\ The vertical and horizontal lines are the $\tau =0$\ and
the past infinity $\tau =-\infty $ slices of the spacetime, respectively and
the double line denotes the future infinity $\tau =+\infty .$ The solid
black dots denote the quasiregular singularities. The region that is outside
the cosmological horizon is located inside the triangle denoted by ``X''. 

Quasiregular singularities are the end points of incomplete and inextendible
geodesics which spiral infinitely around a topologically closed spatial
dimension. Moreover the world lines of observers approaching these points
come to an end in a finite proper time \cite{Konkowski}. They are the
mildest kinds of singularity in that the Riemann tensor and its derivatives
remain finite in all parallelly propagated\ orthonormal frames. Consequently
observers do not see any divergences in physical quantities when they
approach a quasiregular singularity. The flat Kasner spacetimes on the
manifolds $R\times T^{3}$\ or $R^{3}\times S^{1}$, Taub-NUT spacetimes and
Moncrief spacetimes are some typical spacetimes with quasiregular
singularities.\ 

We consider these quasiregular singularities to be quite different from the
cosmological singularities \ referred to in the maximal mass conjecture. \
This conjecture states that a timelike singularity will be present for any
adS spacetime whose conserved mass (\ref{Mcons}) is positive (i.e. larger
than the zero value of de Sitter spacetime). \ Using Schwarzschild de Sitter
spacetime as a paradigmatic example , it is straightforward to show that
scalar Riemann curvature invariants will diverge for $M>0$ \cite%
{Shiro,bala,GM}, yielding a timelike boundary to the manifold upon their
excision. \ Note that such curvature invariants diverge in certain regions
even if $M>0$ ; however observers at future infinity cannot actively probe
such regions. We therefore interpret the cosmological singularities in the
conjecture of ref. \cite{bala} to imply that scalar Riemann curvature
invariants will diverge to form timelike regions of geodesic incompleteness
whenever the conserved mass of a spacetime becomes positive (i.e. larger
than the zero value of pure dS). By this definition quasiregular
singularities are clearly not cosmological singularities, and vice-versa.

\subsection{R-approach}

For simplicity we shall consider the form of the metric when the base space
is a product \ $\otimes _{i=1}^{k}S^{2}$\ of 2-spheres 
\begin{equation}
ds_{R}^{2}=V(\tau )\left( dt+\sum_{i=1}^{k}2n\cos (\theta _{i})d\phi
_{i}\right) ^{2}-\frac{d\tau ^{2}}{V(\tau )}+(\tau
^{2}+n^{2})\sum_{i=1}^{k}(d\theta _{i}^{2}+\sin ^{2}(\theta _{i})d\phi
_{i}^{2})  \label{TBDSgen}
\end{equation}%
keeping in mind that the results below apply to the more general case (\ref%
{TNDSgeneral}) given above. \ We shall denote the largest root of $V(\tau )$
by $\tau _{c}$. \ The subspace for which $\tau =\tau _{c}$ is the fixed
point set of $\partial /\partial t$.

Since $\frac{\partial }{\partial \phi _{1}},\cdot \cdot \cdot ,\frac{
\partial }{\partial \phi _{k}}$\ are Killing vectors, for any constant $
(\phi _{1},\cdot \cdot \cdot ,\phi _{k})$-slice near the horizon $\tau =\tau
_{c}$\ additional conical singularities will be introduced in the $(t,\tau )$
Euclidean section unless $t$\ has period 
\begin{subequations}
\begin{equation}
\beta _{R}=4\pi /\left| V^{\prime }(\tau _{c})\right|  \label{betagen}
\end{equation}%
This periodicity must match the one induced by the requirement that the
Misner string singularities vanish. This yields 
\end{subequations}
\begin{equation}
\frac{1}{\left| V^{\prime }(\tau _{c})\right| }=\frac{(d+1)\left| n\right| }{
2q}  \label{periodicities}
\end{equation}%
which in general has two solutions for $\tau _{c}=\tau ^{\pm }$ as a
function of $n$. For each of these solutions the fixed point set of $
\partial /\partial t$ is $\left( d-1\right) $-dimensional, and so both are
bolt solutions. We shall refer to these solutions as R$^{+}$ and R$^{-}$
respectively, denoting their metric (\ref{TBDSgen}) as $ds_{R}^{2}$.

The general form for the metric determinant $g_{R}$ and the Ricci scalar for
arbitrary dimension $(d+1)$ are 
\begin{eqnarray}
g_{R} &=&-(\tau ^{2}+n^{2})^{(2k)}\prod_{i=1}^{k}\sin ^{2}(\theta _{i})
\label{TBDSdetg} \\
R_{R} &=&\frac{d(d+1)}{\ell ^{2}}  \label{TBDSRs}
\end{eqnarray}%
The bulk action (\ref{actionbulk}) can then be computed for arbitrary $d$
(recalling that the $\prod \sin ^{2}(\theta _{i})$ will contribute to the
volume term) giving 
\begin{equation}
I_{R,B}=\frac{d\beta (4\pi )^{k}}{8\pi \ell ^{2}}\int d\tau \left( \tau
^{2}+n^{2}\right) ^{k}  \label{TBDSBulkgeneral}
\end{equation}%
where ${\frac{4\pi }{|V^{\prime }(\tau _{c})|}=}\beta >0$ is the period of $%
t $.

We work in the region $\tau >\tau _{c}$ near future infinity. Employing the
binomial expansion on the integrand, we integrate term by term from $\tau
=\tau_{c}$ to $\tau \rightarrow \infty $. Since we are only after the finite
contributions (the divergent terms being cancelled by the counter-terms),
the resultant contribution from the bulk action is 
\begin{equation}
I_{R,B\text{finite}}=-\frac{d\beta (4\pi )^{k}}{8\pi \ell ^{2}}
\sum_{i=0}^{k} {\binom{k}{i}} n^{2i}\frac{\tau _{c}^{2k-2i+1}}{ 2k-2i+1}
\label{TBDSBulkfinite}
\end{equation}
Turning now to the boundary contributions at future infinity, the boundary
metric is given by $\gamma _{\mu \nu }=g_{\mu \nu }+n_{\mu }n_{\nu }$, where 
$n_{\mu }=\left[ 0,{\frac{1}{\sqrt{-g^{\tau \tau }}}},0,\ldots \right] $ is
the unit norm of a surface of fixed $\tau $. We obtain for the boundary
metric determinant and its associated Ricci scalar 
\begin{eqnarray}
\gamma _{R} &=&V(\tau )(\tau ^{2}+n^{2})^{2k}\prod_{i=1}^{k}\sin ^{2}(\theta
_{i})  \label{TBDSdgamma} \\
R_{R}(\gamma ) &=&(d-1)\left[ \frac{1}{(\tau ^{2}+n^{2})}-\frac{V(\tau
)n^{2} }{(\tau ^{2}+n^{2})^{2}}\right]  \label{TBDSRgamma}
\end{eqnarray}%
where the trace of the extrinsic curvature can be obtained from the metric (%
\ref{TBDSgen}): 
\begin{equation}
\Theta _{R}=-\left[ \frac{V^{\prime }(\tau )}{2\sqrt{V(\tau )}}+\frac{
(d-1)\tau \sqrt{V(\tau )}}{(\tau ^{2}+n^{2})}\right]  \label{TBDSTrTheta}
\end{equation}

Expanding (\ref{TBDSdgamma}) and (\ref{TBDSTrTheta}) in (\ref{actionboundary}%
) for large $\tau$, the finite contribution from the boundary action will be 
\begin{equation}
I_{R,\partial B \text{finite}} = - \frac{ \beta (4\pi)^k d }{8\pi } m
\label{TBDSboundaryfinite}
\end{equation}

Turning next to the counter-term contributions, which can be found from (\ref%
{TBDSdgamma}) and (\ref{TBDSRgamma}), it can be shown, using exactly the
same arguments employed in \cite{CFM}, that only the first term in the
counter-term action (\ref{actionct}) contributes a finite term - all the
other terms will only cancel the divergences in the bulk and boundary
actions. Hence, the finite contribution at future infinity from the
counter-term action is 
\begin{equation}
I_{R,ct\text{finite}}=\frac{\beta (4\pi )^{k}(d-1)}{8\pi }m
\label{TBDSctfinite}
\end{equation}%
Adding together (\ref{TBDSBulkfinite}), (\ref{TBDSboundaryfinite}) and (\ref%
{TBDSctfinite}), we find 
\begin{equation}
I_{R(\text{finite})}=-\frac{\beta (4\pi )^{k}}{8\pi }\left[ m+\frac{d}{\ell
^{2}}\sum_{i=0}^{k}{\binom{k}{i}}n^{2i}\frac{\tau _{c}^{2k-2i+1}}{(2k-2i+1)}%
\right]  \label{TBDSItotgen}
\end{equation}%
for the general form of the R-approach action.

We now turn to an evaluation of the conserved charges from the formula 
\begin{equation}
\mathfrak{Q}=\oint d^{d-1}x~\sqrt{\gamma }~T_{ab}n^{a}\xi ^{b}  \label{Qgen}
\end{equation}
The only non-vanishing conserved charge will be the conserved mass
associated with $\xi =\partial _{t}$. Thus we have 
\begin{equation}
\mathfrak{M}=\frac{1}{8\pi }\int d^{d-1}x~\sqrt{\gamma }\left\{ \Theta
_{ab}-\Theta \gamma _{ab}+\frac{(d-1)}{\ell }\gamma _{ab}+\ldots \right\}
n^{a}\xi ^{b}  \label{Mgen}
\end{equation}
The extra terms from the variation of the counter-term action can be found
in \cite{GM}. Using exactly the same arguments as above (from \cite{CFM}),
it can be shown that only the first term ${\textstyle\frac{(d-1)}{\ell }}
\gamma_{ab}$ contributes to the finite conserved mass. Inserting all of the
quantities, we find that the finite conserved mass for general $(d+1)=2k+2$
dimensions is given by 
\begin{equation}
\mathfrak{M}_{R}=-\frac{(4\pi )^{k}2k}{8\pi }m  \label{Massgeneral}
\end{equation}
$m$ can be solved for in terms of $\tau ,n$, through the first condition of
demanding that $V(\tau )=0$.

Using (\ref{Massgeneral}), (\ref{TBDSItotgen}), and the Gibbs-Duhem relation
(\ref{GDoutfinal}), we obtain 
\begin{equation}
S_{R}=\frac{(4\pi )^{k}\beta }{8\pi }\left\{ \frac{d}{\ell ^{2}}%
\sum_{i=0}^{k}{\binom{k}{i}}n^{2i}\frac{\tau _{c}^{2k-2i+1}}{2k-2i+1}%
-(2k-1)m\right\}  \label{TBDSStotgen}
\end{equation}%
as the expression for the entropy for the Taub-Bolt-dS spacetime in general
dimension $(d+1)=2k+2$. \ 

Note that none of the preceding results required imposition of the
consistency condition (\ref{periodicities}), which also reads 
\begin{equation}
|V^{\prime }(\tau _{c})|=\frac{2q}{\left( d+1\right) \left| n\right| }
\label{periodreq}
\end{equation}
\ Eq. (\ref{periodreq}) has in general four solutions for $\tau _{c}$, two
of which are positive, yielding two possible relationships between the
parameters $m$ and $n$. This in turn implies two distinct spacetimes, each
with its own characteristic entropy and conserved mass for a given $n$.
While eq. (\ref{periodreq}) is easily solvable for specific choice of $d$,
it is cumbersome to solve for arbitrary $d$, and so we shall postpone
analysis of the implementation of this condition.

\subsection{C-approach}

The form of the metric in this approach is obtained from (\ref{TBDSgen}) by
rotating the time and the NUT parameter ($t\rightarrow iT,~n\rightarrow iN$%
), giving 
\begin{equation}
ds_{C}^{2}=-F(\rho ) \left( dT + \sum_{i=1}^k 2N\cos (\theta _{i})d\phi _{i}
\right) ^{2} - \frac{d\rho ^{2}}{F(\rho )}+(\rho
^{2}-N^{2})\sum_{i=0}^{k}(d\theta _{i}^{2}+\sin ^{2}(\theta _{i})d\phi
_{i}^{2})  \label{TNDSgen}
\end{equation}%
where $F(\rho )$ is now given by 
\begin{equation}
F(\rho )=\frac{2m\rho }{(\rho ^{2}-N^{2})^{k}}-\frac{\rho }{(\rho
^{2}-N^{2})^{k}}\int_{\rho }ds\left[ \frac{(s^{2}-N^{2})^{k}}{s^{2}}-\frac{%
(2k+1)}{\ell ^{2}}\frac{(s^{2}-N^{2})^{k+1}}{s^{2}}\right]  \label{FtNUTgen}
\end{equation}

Since these two formulae are the same as in the Bolt case, except for a few
signs, the same arguments used above can be used to find the finite action
and entropy, as well as the conserved mass. The general metric determinant
and Ricci scalar are 
\begin{eqnarray}
g_{C} &=&(\rho ^{2}-N^{2})\prod_{i=1}^{k}\sin (\theta _{i})  \label{TNDSdetg}
\\
R_{C} &=&\frac{d(d+1)}{\ell ^{2}}  \label{TNDSRs}
\end{eqnarray}%
The finite contribution to the bulk action can again be found by inserting
the above into (\ref{actionbulk}) and using the binomial expansion, 
\begin{equation}
I_{C,B\text{finite}}=-\frac{(4\pi )^{k}\beta d}{8\pi \ell ^{2}}%
\sum_{i=0}^{k} {\binom{k}{i}} (-1)^{i}N^{2i}\frac{\rho _{+}^{2k-2i+1}}{%
2k-2i+1}  \label{TNDSBulkgeneral}
\end{equation}%
with $\rho_{+}$ the largest positive root of $F(\rho )$, found by the fixed
point set of $\partial _{T}$.

The quantity $\beta $ is the period of $T$, again obtained by setting 
\begin{equation*}
\beta_C ={\frac{4\pi }{|F^{\prime }(\rho_{+})|}=}\frac{2(d+1)\pi \left|
N\right| }{q}
\end{equation*}
so as to ensure regularity in the $\left( T,\rho \right) $ section. Note
that in this approach the functional form of $F(\rho )$ is altered due to
the changes in signs, so that $\rho_{+}$ is not equal to $\tau _{c}$

\bigskip

The boundary metric is again given by $\gamma _{\mu \nu }=g_{\mu \nu
}+n_{\mu }n_{\nu }$, with $n_{\mu }$ as before, and again we work at future
infinity. This gives 
\begin{eqnarray}
\gamma _{C} &=&-F(\rho )(\rho^{2}-N^{2})^{2k}\prod_{i=1}^{k}\sin ^{2}(\theta
_{i})  \label{TNDSgamma} \\
R_{C}(\gamma ) &=&(d-2)\left[ \frac{1}{(\rho ^{2}-N^{2})}+\frac{F(\rho
)N^{2} }{(\rho ^{2}-N^{2})^{2}}\right]  \label{TNDSRsgamma}
\end{eqnarray}%
for the boundary metric determinant and Ricci scalar. The trace of the
extrinsic curvature on the boundary can be found from (\ref{TNDSgen}) 
\begin{equation}
\Theta_{C}=-\left[ \frac{F^{\prime }(\rho )}{2\sqrt{F(\rho )}}+\frac{
(d-1)\rho \sqrt{F(\rho )}}{(\rho ^{2}-N^{2})}\right]  \label{TNDSTheta}
\end{equation}
Using the same steps as above, the finite contributions from the boundary
and counter-term actions can be found to be the same as (\ref%
{TBDSboundaryfinite}), (\ref{TBDSctfinite}), and so, for the Taub-NUT-dS
metric, the general action is calculated to be 
\begin{equation}
I_{C~\text{finite}}=-\frac{(4\pi )^{k}\beta }{8\pi }\left[ m+\frac{d}{\ell
^{2}}\sum_{i=0}^{k}{\binom{k}{i}}(-1)^{i}N^{2i}\frac{\rho_{+}^{2k-2i+1}}{
2k-2i+1}\right]  \label{TNDSactiongeneral}
\end{equation}%
As in the R-approach, the conserved mass can be found from (\ref{Qgen}),
using the expansion (\ref{Massgeneral}). Again, only the three terms given
in (\ref{Massgeneral}) contribute to the finite conserved mass, giving for
the C-approach spacetimes 
\begin{equation}
\mathfrak{M}_{C}=-\frac{(4\pi )^{k}2k}{8\pi }m  \label{TNDSGenMass}
\end{equation}%
Using once more the relation $S=\beta H_{\infty }-I$, the expression for the
entropy for the general Taub-NUT-dS spacetime is 
\begin{equation}
S_{C}=\frac{(4\pi )^{k}\beta }{8\pi }\left[ \frac{d}{\ell ^{2}}\sum_{i=0}^{k}%
{\binom{k}{i}}(-1)^{i}N^{2i}\frac{\rho _{+}^{2k-2i+1}}{2k-2i+1}-(2k-1)m%
\right]  \label{TNDSentropygen}
\end{equation}

\bigskip

The periodicity conditions for ensuring regularity in the $\left( T,\tau
\right) $ section and removal of all Misner-string singularities now yields
the consistency requirement 
\begin{equation}
|F^{\prime }(\rho _{+})|=\frac{2q}{\left( d+1\right) \left| N\right| }
\label{periodicNUT}
\end{equation}%
In this case there are two qualitatively distinct solution classes to (\ref%
{periodicNUT}), characterized by the co-dimensionality of the fixed point
set of $\partial _{T}$.\ \ In one class, this co-dimensionality is $\left(
d-1\right) $, yielding a solution $\rho _{+}>N$ -- we shall refer to this as
the Taub-Bolt-C solution. In the second class $\rho _{+}=N$, and the
fixed-point set is of zero-dimensionality. This class shall be referred to
as the Taub-NUT-C solution. These different cases will be treated in more
detail in specific dimensions in the sequel.

\section{Four Dimensional Analysis}

\label{sec:4d}

\subsection{R-approach in 4 dimensions}

The ($3+1$)-dimensional metric will in this case ((\ref{TBDSgen}), with $k=1$%
) have 
\begin{equation}
V(\tau )=\frac{\tau ^{4}+(6n^{2}-\ell ^{2})\tau ^{2}+n^{2}(\ell
^{2}-3n^{2})+2m\tau \ell ^{2}}{(\tau ^{2}+n^{2})\ell ^{2}}  \label{TB4}
\end{equation}%
where $n$ is the non-vanishing NUT charge and $\Lambda ={\frac{3}{\ell ^{2}}}
$. The coordinate $t$ parameterizes a circle fibered over the 2-sphere with
coordinates $\left( \theta ,~\phi \right) $, and must have a period
respecting the condition (\ref{periodreq}), which is 
\begin{equation}
\beta _{R}=\frac{4\pi }{|V^{\prime }(\tau _{c})|}=\frac{8\pi \left| n\right| 
}{q}  \label{pTB4}
\end{equation}%
where $q$ is a positive integer, yielding 
\begin{equation}
\beta _{R}=2\pi \left| \frac{(\tau _{c}^{2}+n^{2})^{2}\ell ^{2}}{-2\tau
_{c}\ell ^{2}n^{2}+\tau _{c}^{5}+2\tau _{c}^{3}n^{2}+9n^{4}\tau _{c}-m\ell
^{2}\tau _{c}^{2}+m\ell ^{2}n^{2}}\right|  \label{TBDSbet4d}
\end{equation}%
The geometry of a constant-$\tau $ surface is that of a Hopf fibration of $%
S^{1}$ over $S^{2}$, and the metric (\ref{TB4}) describes the
contraction/expansion (for $q=1$) of this 3-sphere in spacetime regions
where $V(\tau )>0$ outside of the past/future cosmological horizons. The
condition $V(\tau _{c})=0$ yields 
\begin{equation}
m_{R}=-\frac{\tau _{c}^{4}-\ell ^{2}\tau _{c}^{2}+n^{2}\ell ^{2}+6n^{2}\tau
_{c}^{2}-3n^{4}}{2\ell ^{2}\tau _{c}}  \label{TBDSmb4d}
\end{equation}

Using the general formula (\ref{TBDSItotgen}) with $k=1,~d=3$, the action is 
\begin{equation}
I_{R,4d}=-\frac{\beta }{2\ell ^{2}}(m_{R}\ell ^{2}+\tau _{c}^{3}+3n^{2}\tau
_{c})  \label{TBDS4dItot}
\end{equation}
and the conserved mass is found to be 
\begin{equation}
\mathfrak{M}_{R,4d}=-m_{R}+\frac{\ell ^{4}-30n^{2}\ell ^{2}+105n^{4}}{8\tau
\ell ^{2}}+\mathcal{O}\left( \frac{1}{\tau ^{2}}\right)  \label{TBDSMass4d}
\end{equation}%
near future infinity, which for $n=0$ reduces to the total mass of the four
dimensional Schwarzschild-dS black hole \cite{GM}. From (\ref{TBDSStotgen})
or by directly applying the Gibbs-Duhem relation (\ref{GDoutfinal}) $S=\beta
H_{\infty }-I$, we find 
\begin{equation}
S_{R4d}=-\frac{\beta (m_{R}\ell ^{2}-3n^{2}\tau _{c}-\tau _{c}^{3})}{2\ell
^{2}}  \label{TBDSStot4d}
\end{equation}%
\bigskip for the total entropy.

The preceding results are generic to either of the two solutions $\tau
_{c}=\tau _{c}^{\pm }$ to (\ref{periodreq}), which are 
\begin{equation}
\tau _{c}^{\pm }=\frac{q\ell ^{2}\pm \sqrt{q^{2}\ell
^{4}-144n^{4}+48n^{2}\ell ^{2}}}{12n}  \label{TBDStpm4d}
\end{equation}%
Since the discriminant of $\tau _{c}^{\pm }$ must always be positive, we
find 
\begin{equation}
\left| n_{max}\right| <\frac{\ell }{6}\sqrt{6+3\sqrt{4+q^{2}}}
\label{TBDSnmax4d}
\end{equation}
Note that both the high temperature ($n\rightarrow 0$) and flat space ($\ell
\rightarrow \infty $) limits of $\tau _{c}^{+}$ are infinite. The high
temperature limit of $\tau _{c}^{-}$ is $0$, and its flat space limit is $- 
\frac{2n}{q}$. From these results we have mass and temperature parameters $
\beta ^{\pm }$ and $m^{\pm }$ , straightforwardly obtained by insertion of $
\tau _{c}=\tau _{c}^{\pm }$ into eqs. (\ref{TBDSbet4d}) and (\ref{TBDSMass4d}%
) respectively. We shall refer to the distinct spacetimes associated with
these cases as R$_{4}^{\pm }$, with action $I_4^{\pm }$ and entropy $%
S_4^{\pm }$ .

\begin{figure}[tbp]
\centering        
\begin{minipage}[c]{.45\textwidth}
         \centering
         \includegraphics[width=\textwidth]{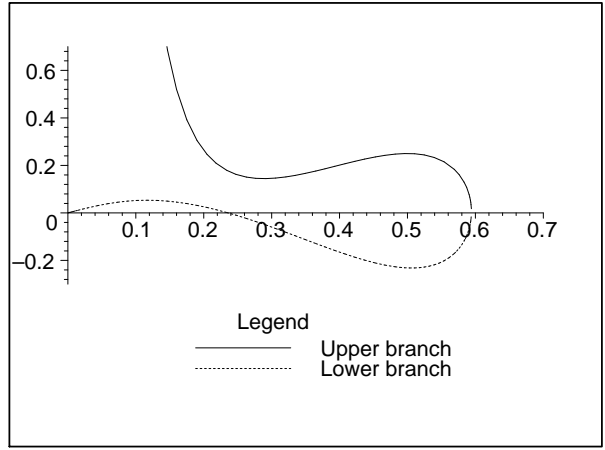}
         \label{PlotTBDSmbpm4d}
     \end{minipage}
\caption{Plot of the upper ($\protect\tau _{b}=\protect\tau _{b+}$) and
lower ( $\protect\tau _{b}=\protect\tau _{b-}$) TB masses (for $q=1$).}
\end{figure}

Further analysis indicates that the R$_{4}^{\pm }$ spacetimes provide a
counter-example to the maximal mass conjecture of \cite{bala} for certain
ranges of the parameter $n$. As shown in Figure \ref{PlotTBDSmbpm4d} (with $%
q=1$), the $\mathfrak{M}_{R}^{+}$ is always positive, and thus always
violates the conjecture. Note that although the $\mathfrak{M}_{R}^{-}$ is
positive for $n<.2360026142\ell $, it doesn't violate the conjecture for $n$
\ less than this value, since R$^{-}$\ exists only for $\left| n\right|
>.2658\,\ell $.\ Otherwise $V(\tau )$\ develops two additional larger real
roots, and the periodicity condition cannot be satisfied. Note that for $q>1$%
, the lower branch R$^{-}$\ always has a negative $\mathfrak{M}_{R}^{-}$ and
so does not violate the conjecture, whereas the R$^{+}$ branch violates the
conjecture for all $q$, since $\mathfrak{M}_{R}^{+}>0$.

\begin{figure}[tbp]
\centering       
\begin{minipage}[c]{.45\textwidth}
         \centering
         \includegraphics[width=\textwidth]{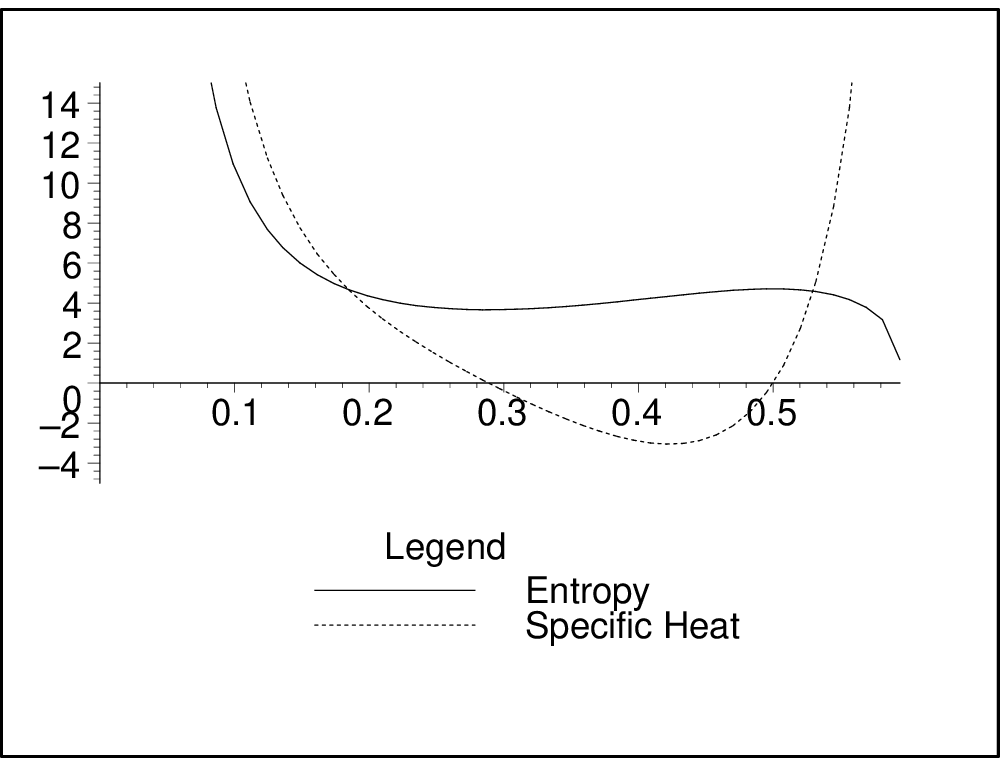}
         \caption{Plot of the upper branch bolt entropy and specific heat (for $q=1$).}
         \label{PlotSCTBp4d}
\end{minipage}\begin{minipage}[c]{0.05\textwidth}
\end{minipage}%
\begin{minipage}[c]{.45\textwidth}
         \centering
         \includegraphics[width=\textwidth]{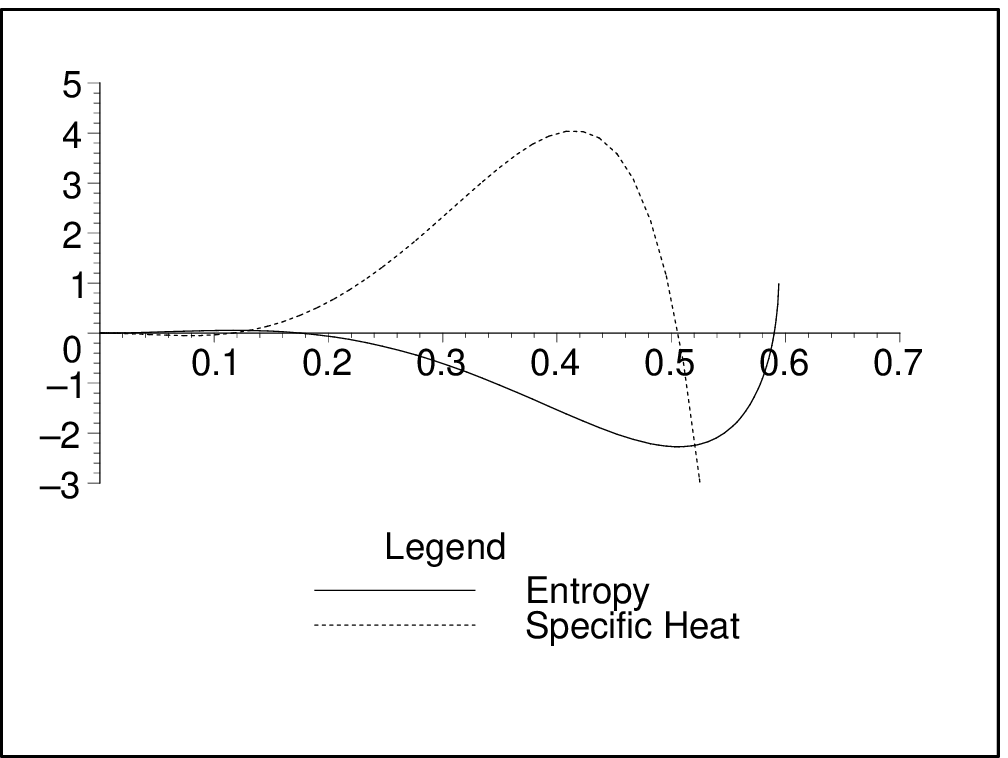}
         \caption{Plot of the lower branch bolt entropy and specific heat (for $q=1$).}
         \label{PlotSCTBm4d}
\end{minipage}
\end{figure}

From (\ref{TBDS4dItot}), using (\ref{TBDSbet4d}) and (\ref{TBDStpm4d}), the
R action is 
\begin{eqnarray}
I_{R,4d}^{\pm }(\tau _{0}=\tau _{0}^{\pm }) &=&-\frac{\pi \ell ^{2}}{216}%
\frac{(72n^{2}+q^{2}\ell ^{2})}{n^{2}}  \label{TBDSIpm4d} \\
&&\pm \frac{\pi }{216}\frac{(-q^{2}\ell ^{4}+144n^{4}-48n^{2}\ell ^{2})\sqrt{%
q^{2}\ell ^{4}-144n^{4}+48n^{2}\ell ^{2}}}{n^{2}q\ell ^{2}}  \notag
\end{eqnarray}%
and from (\ref{TBDSStot4d}), the entropy is 
\begin{eqnarray}
S_{R,4d}^{\pm } &=&\frac{\pi \ell ^{2}(24n^{2}+q^{2}\ell ^{2})}{72n^{2}}
\label{TBDSSpm4d} \\
&&\pm \frac{\pi (144n^{4}+q^{2}\ell ^{4})\sqrt{q^{2}\ell
^{4}-144n^{4}+48n^{2}\ell ^{2}}}{72n^{2}q\ell ^{2}}  \notag
\end{eqnarray}%
This does satisfy the first law, though each branch must be checked
separately. From the entropy and the relation $C_{R,4d}^{\pm }=-\beta
_{R}^{\pm }\partial _{\beta _{R}^{\pm }}S_{R,4d}^{\pm }$, we find for the
specific heat 
\begin{equation}
C_{R}^{\pm }(\tau _{0}^{\pm })=\frac{\pi \ell ^{4}q^{2}}{36n^{2}}\pm \frac{%
\pi (-144q^{2}\ell ^{4}n^{4}+41472n^{8}-10368n^{6}\ell ^{2}+24n^{2}\ell
^{6}k^{2}+k^{4}\ell ^{8})}{q\ell ^{2}n^{2}\sqrt{q^{2}\ell
^{4}-144n^{4}+48n^{2}\ell ^{2}}}  \label{TBDSspeheat}
\end{equation}%
The plots for the upper and lower branch entropies/specific heats are in
Figures \ref{PlotSCTBp4d}, \ref{PlotSCTBm4d}. 

In figure \ref{PlotSCTBp4d}, the upper branch entropy is always positive
(and almost always greater than $\pi \ell ^{2},$ except for NUT charge in a
range near the maximal value $n_{max}=.5941\ell $), but the specific heat is
positive only outside the range $.2886751346\ell <n<.5\ell $; thus, the
upper branch solutions are only stable for $n$ outside this range. Figure %
\ref{PlotSCTBm4d} shows that the lower branch entropy is always negative,
and so the lower branch solutions are always unstable.

\begin{figure}[tbp]
\centering        
\begin{minipage}[c]{.4\textwidth}
         \centering
         \includegraphics[width=\textwidth]{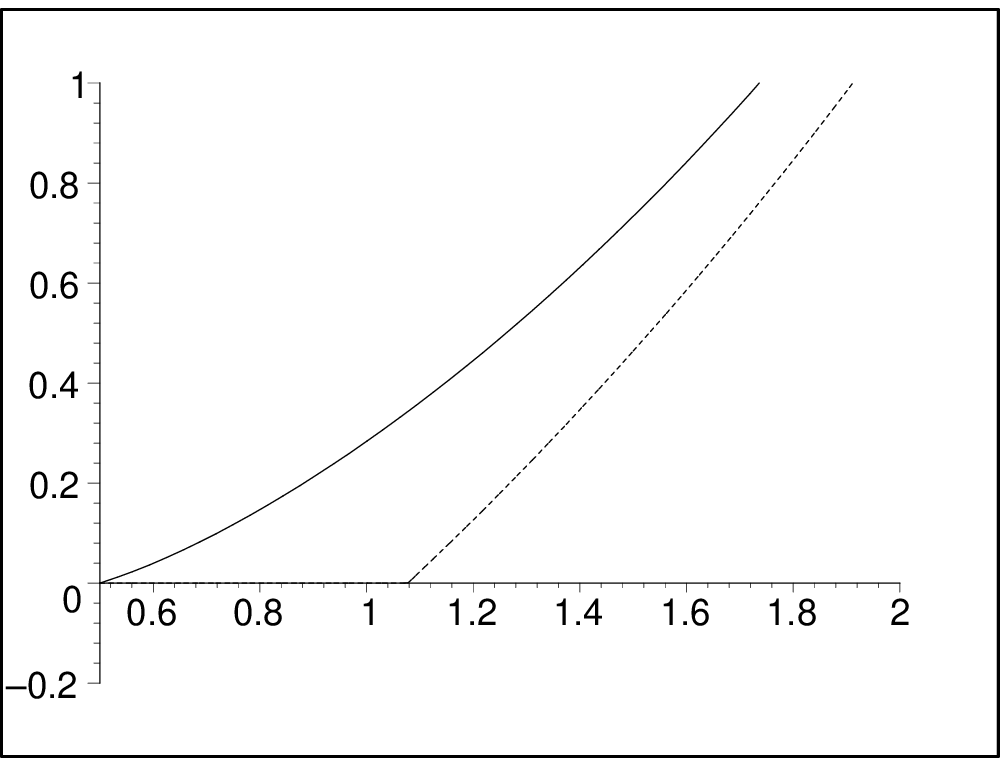}
         \caption{$c$-function of Taub-Bolt-dS solution versus
$\tau \geq \tau ^{+}=0.50$ for $q=1$ (solid) and $\tau \geq \tau ^{+}=1.077$
for $q=3$ (dotted). }
         \label{CPLUS}
\end{minipage}
\begin{minipage}[c]{0.05\textwidth}
\end{minipage}
\begin{minipage}[c]{.4\textwidth}
         \centering
         \includegraphics[width=\textwidth]{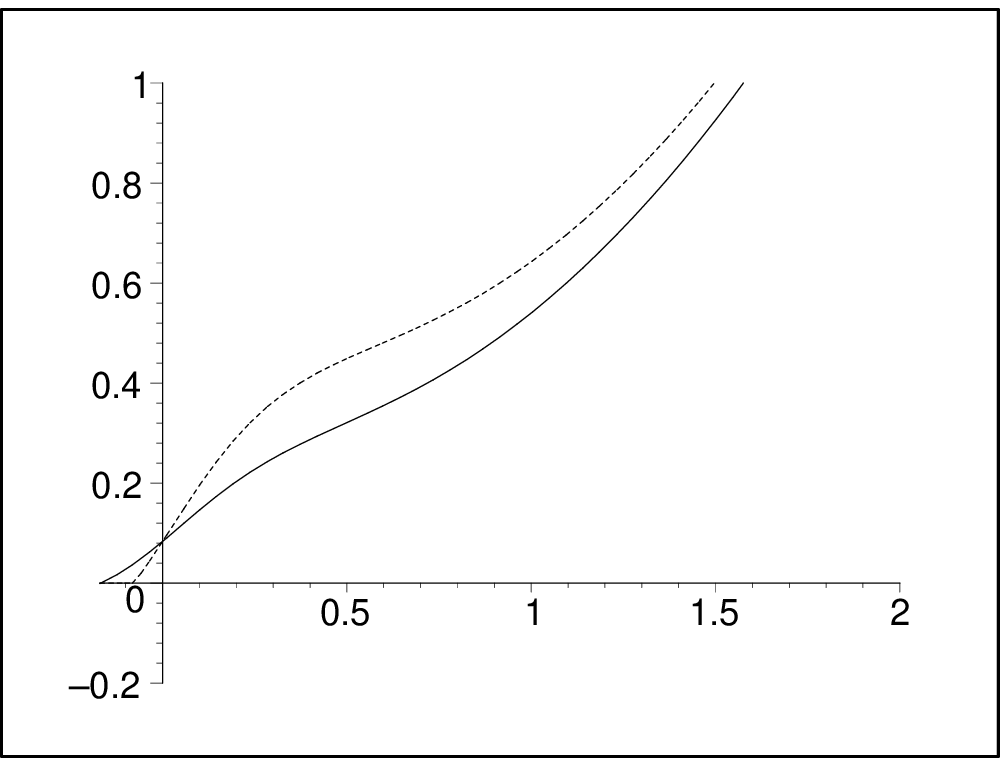}
         \caption{$c$-function of Taub-Bolt-dS solution versus
$\tau \geq \tau ^{-}=-0.167$ for $q=1$ (solid) and $\tau \geq \tau ^{-}=-0.077$
for $q=3$ (dotted).}
         \label{CMINUS}
\end{minipage}\label{fig55}
\end{figure}

In both the AdS and dS cases there is a natural correspondence between
phenomena occurring near the boundary (or in the deep interior) of either
spacetime and UV (IR) physics in the dual CFT. Solutions that are
asymptotically (locally) dS lead to an interpretation in terms of
renormalization group flows and an associated generalized dS $c$-theorem.
This theorem states that in a contracting patch of dS spacetime, the
renormalization group flows toward the infrared and in an expanding
spacetime, it flows toward the ultraviolet. Since the spacetime (\ref%
{TBDSgen}) is asymptotically (locally) dS, we can use the four-dimensional $%
c $-function \cite{Leb} 
\begin{equation}
c=\left( G_{\mu \nu }n^{\mu }n^{\nu }\right) ^{-1}=\frac{1}{G_{\tau \tau }}
\label{cfunction}
\end{equation}%
where $n^{\mu }$\ is the unit normal vector to a constant $\tau -$slice. In
figures (\ref{CPLUS}) and (\ref{CMINUS}), the diagrams of the R-approach
spacetime $c$-functions outside the cosmological horizon with $\ell =1$ and $%
n=0.5$ for two cases $q=1$ and $3$ are plotted. 

As one can see from these figures, outside the cosmological horizon, the $c$
-function is a monotonically increasing function of $\ \tau ,$ indicative of
the expansion of a constant $\tau $-surface of the metric (\ref{TBDSgen})
outside of the cosmological horizon$.$ Since the metric (\ref{TBDSgen}) at
future infinity $\tau \rightarrow +\infty ,$ reduces to 
\begin{equation}
ds_{R}^{2}\rightarrow -du^{2}+e^{2u/\ell }d\Sigma _{3}^{2}  \label{TNDSbou}
\end{equation}%
where $u=\ell \ln \tau $ and $d\Sigma _{3}^{2}$ is the metric of
three-dimensional constant $u$-surface, the scale factor in (\ref{TNDSbou})
expands exponentially near future infinity. \ Hence the behavior of $c$%
-function in figures (\ref{CPLUS}) and (\ref{CMINUS}) is in good agreement
with what one expects from the $c$-theorem. According to $c$-theorem for any
asymptotically (locally) dS spacetimes, the $c$-function must increase
(decrease) for any expanding (contracting) patch of the spacetime.

\textbf{\bigskip }

\subsection{C-approach in 4 dimensions}

The ($3+1$) dimensional C-approach ($t\rightarrow iT$, $n\rightarrow iN$)
implies that the metric has the form (\ref{TNDSgen}), with $k=1,~d=3$, with $%
F(\rho )$ given by 
\begin{equation}
F(\rho )=\frac{\rho ^{4}-(\ell ^{2}+6N^{2})\rho ^{2}+2m\rho \ell
^{2}-N^{2}(\ell ^{2}+3N^{2})}{(\rho ^{2}-N^{2})\ell ^{2}}  \label{TNDSFr4d}
\end{equation}%
where $N$ is the nonvanishing NUT charge and $\Lambda ={\textstyle\frac{3}{%
\ell ^{2}}}$. The coordinate $T$ parameterizes a circle fibered over the
non-vanishing sphere parameterized by $(\theta ,\phi )$ and must have
periodicity respecting the following condition 
\begin{equation}
\beta _{C}=\frac{4\pi }{|F^{\prime }(\rho )|}=\frac{8\pi |N|}{q}
\label{TNDS4dbeta}
\end{equation}%
to avoid conical singularities, where $q$ is a positive integer.

For this situation (i.e. with $\left( --++\right) $\ signature) the geometry
is, strictly speaking, no longer that of a Hopf fibration of $S^{1}$\ over a
2-sphere since the coordinate $T$ is now timelike. \ Consequently its
physical relevance is less clear. However the metric is independent of the
coordinate $T$ and so we are able to proceed to calculate the action, the
conserved mass and various other quantities. We shall do so, mindful of the
preceding considerations.

Using the method of counter-terms for de Sitter space \cite{GM} directly, or
using the general formula (\ref{TNDSactiongeneral}) obtained above with $k=1$
and $d=3$, we find 
\begin{equation}
I_{C,4d}=-\frac{\beta _{C}(\rho _{+}^{3}-3N^{2}\rho _{+}+m\ell ^{2})}{2\ell
^{2}}  \label{TNDS4dItot}
\end{equation}%
for the action in four dimensions, where $\rho _{+}$ is the value of $\rho $
that is the largest positive root of $F(\rho )$, determined by the fixed
point set of $\partial _{T}$, and $m$ is the mass parameter, which will have
differing values for different $\rho \geq \rho _{+}$.

Working at future infinity, using either (\ref{TNDSGenMass}) or calculating
directly from (\ref{Qgen}),(\ref{Mgen}) for the metric (\ref{TNDSgen})
yields the conserved mass 
\begin{equation}
\mathfrak{M}_{C,4d}=-m + \frac{105N^{4} + 30N^{2} \ell^{2} + \ell^{4}}{
8\ell^{2} \rho } +O\left(\frac{1}{\rho^{2}} \right)  \label{TNDS4dtotmass}
\end{equation}
near future infinity, which for $N=0$ reduces exactly to the total mass of
the four dimensional Schwarzschild-dS black hole \cite{GM}.

By applying the Gibbs-Duhem relation (\ref{GDoutfinal}) or from (\ref%
{TNDSentropygen}) we obtain 
\begin{equation}
S_{C,4d}=\frac{\beta _{H}(\rho _{+}^{3}-3N^{2}\rho _{+}-m\ell ^{2})}{2\ell
^{2}}  \label{TNDS4dStot}
\end{equation}
for the total entropy. This entropy can be shown to satisfy the first law of
gravitational thermodynamics (as required) for both the NUT and bolt cases
(see below). The above equations are generic, and can now be analyzed for
the specific cases of the ``NUT'' and ``Bolt'' solutions (called such in
analogue with the Taub-NUT-AdS case \cite{CFM}).

The metric arising from the C-approach affords two sets of solutions,
depending on the fixed point set of $\partial _{T}$. These arise from the
regularity condition (\ref{TNDS4dbeta}) that ensures the absence of conical
singularities. When $\rho _{+}=N$, $F(\rho =N)=0$ and the fixed point set of 
$\partial _{T}$ is $0$-dimensional, we get the ``NUT'' solutions; when $\rho
_{+}=\rho _{b\pm }>N$, the fixed point set is 2-dimensional, we get the
``bolt'' solutions. Since the thermodynamic and mass analysis yield
interesting yet different results for each case we will handle each
separately.

\subsubsection{Taub-NUT-C Solution}

For the NUT solution, $\rho _{+}=N$, and we can solve for the NUT mass
parameter 
\begin{equation}
m_{C,n}=\frac{N(\ell ^{2}+4N^{2})}{\ell ^{2}}  \label{TNDS4dmn}
\end{equation}%
Looking at this equation, it is easily seen that $m_{C,n}$ is always
positive, and so (since the conserved mass at future infinity is $-m_{C,n}$ (%
\ref{TNDS4dtotmass})) the NUT solution always has a mass less than the
de-Sitter mass, always satisfying the Balasubramanian et. al. conjecture %
\cite{bala}. In the flat space limit ($\ell \rightarrow \infty $), the NUT
mass will go to $N$, and in the high temperature ($N\rightarrow 0$) limit,
it goes to 0.

The period in four dimensions ($q=1$) is given by $\beta =8\pi N$, and so
the NUT action and entropy can be found from (\ref{TNDS4dItot}, \ref%
{TNDS4dStot}); 
\begin{eqnarray}
I_{C,NUT4d} &=&-\frac{4\pi N^{2}(\ell ^{2}+2N^{2})}{\ell ^{2}}
\label{TNDS4dINUT} \\
S_{C,NUT4d} &=&-\frac{4\pi N^{2}(\ell ^{2}+6N^{2})}{\ell ^{2}}
\label{TNDS4dSNUT}
\end{eqnarray}%
It is easy to show that (\ref{TNDS4dSNUT}) and (\ref{TNDS4dtotmass}) with $
m=m_{C,n}$ satisfy the first law $dS=\beta ~dH$. In the flat space limit,
both of these go to $-4\pi N$, and in the high temperature limit, they both
go to $0$.

Using (\ref{TNDS4dSNUT}) and the relation $C=-\beta \partial _{\beta }S$
yields 
\begin{equation}
C_{C,NUT4d}=\frac{8\pi N^{2}(\ell ^{2}+12N^{2})}{\ell ^{2}}  \label{CNUT}
\end{equation}%
for the NUT specific heat. \ In the flat space limit, $C_{C,NUT4d}
\rightarrow 8\pi N$, and it approaches $0$ in the high temperature limit.

We note that the specific heat is seen to be always positive, and the
entropy is always negative for the NUT solution. We interpret this to mean
that the NUT solution is not thermodynamically stable (see Figure \ref%
{PlotSCNUT}). Also, the specific heat always negative means it is always
less than the pure dS entropy, thus satisfying the N-bound.

\begin{figure}[tbp]
\centering        
\begin{minipage}[c]{.45\textwidth}
         \centering
         \includegraphics[width=\textwidth]{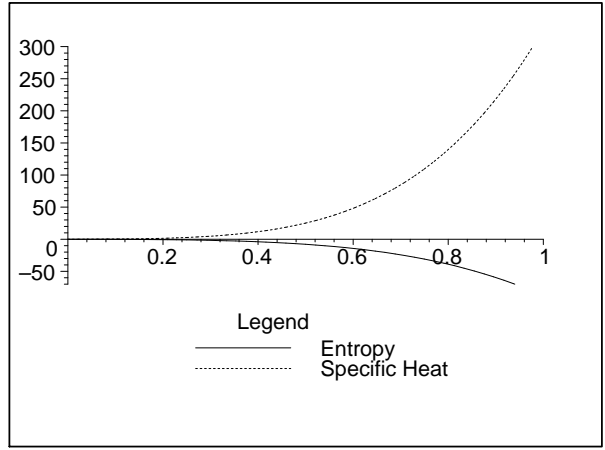}
    \end{minipage}
\caption{Plot of the NUT entropy and specific heat vs. $N.$}
\label{PlotSCNUT}
\end{figure}

\subsubsection{Taub-Bolt-C Solution}

For the bolt solution, the fixed point set of $\partial _{T}$ is 2
dimensional, and we get $\rho _{+}=\rho _{b\pm }>N$. The conditions for a
regular bolt solution are (i) $F(\rho )=0$ and (ii) $F^{\prime }(\rho )=\pm {%
\ \ \textstyle\frac{q}{2N}}$, with (ii) arising from the second equality in (%
\ref{TNDS4dbeta}) (and $N>0$). From (i), we get the bolt mass parameter 
\begin{equation}
m_{C,b}=-\frac{(\rho _{b}^{4}-(\ell ^{2}+6N^{2})\rho _{b}^{2}-N^{2}(\ell
^{2}+3N^{2}))}{2\ell ^{2}\rho _{b}}  \label{TNDS4dmb}
\end{equation}%
where from (ii$^{+}$), $\rho _{b}$ is 
\begin{equation}
\rho _{b\pm }=\frac{q\ell ^{2}\pm \sqrt{q^{2}\ell ^{4}+48N^{2}\ell
^{2}+144N^{4}}}{12N}  \label{TNDS4dtbpm}
\end{equation}%
The discriminant of $\rho _{b\pm }$ will always be positive, and so there is
no restriction on the range of $N$ (except $N>0$). Note that both the flat
space and high temperature limits of $\rho _{b+}$ are infinite; the flat
space limit of $\rho _{b-}$ is $-{\textstyle\frac{2N}{k}}$, and the high
temperature limit is 0.

The period for the bolt is found from the first equality in (\ref{TNDS4dbeta}%
) 
\begin{equation}
\beta_{C,bolt4d} = 2 \pi \Bigg| \frac{ (\rho_b^2 - N^2 )^2 \ell^2 }{
\rho_b^5 - 2 N^2 \rho_b^3 + N^2 ( 9 N^2 + 2 \ell^2 ) \rho_b - m \ell^2 (
\rho_b^2 + N^2 ) } \Bigg|  \label{TNDS4dbetbolt}
\end{equation}
The temperature for the two solutions is the same, as can be seen by
substituting $m=m_{C,b}$ and either of $\rho_b = \rho_{b\pm}$ into (\ref%
{TNDS4dbetbolt}).

\begin{figure}[tbp]
\centering        
\begin{minipage}[c]{.45\textwidth}
        \centering
        \includegraphics[width=\textwidth]{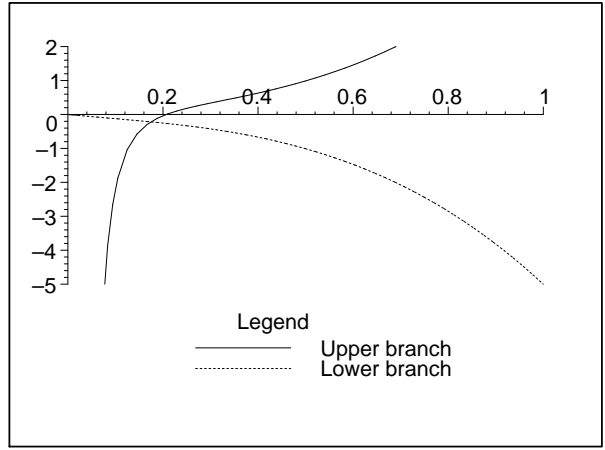}
    \end{minipage}
\caption{Plot of the upper ($\protect\rho _{b}=\protect\rho _{b+}$) and
lower ( $\protect\rho _{b}=\protect\rho _{b-}$) bolt masses $m_{b\pm }$ (for 
$q=1$).}
\label{Plotmbpm4d}
\end{figure}

Substituting in $\rho _{b}=\rho _{b\pm }$ into $m_{C,b}$, we can see that
the Taub-Bolt-C solution is a counter-example to the maximal mass conjecture
of \cite{bala} for certain values of $N$. As shown in Figure \ref{Plotmbpm4d}
(where $q=1$ in the plots), the lower branch ($\rho _{b}=\rho _{b-}$) mass
is always negative, and since (\ref{TNDS4dtotmass}) is $-m$, the lower
branch bolt conserved mass will always be positive, and thus greater than
the de-Sitter mass, violating the conjecture. Also, the upper branch ($\rho
_{b}=\rho _{b+}$) is negative for $N<0.2066200733$, and thus the upper
branch solution also violates the conjecture for $N$ less than this value.
(This trend holds for higher values of $q$, with the cross-over point for
the upper branch solution increasing with increasing $q$).

The bolt action is, using (\ref{TNDS4dItot}) and (\ref{TNDS4dbetbolt}) 
\begin{eqnarray}
I_{C,bolt4d}(\rho _{b}=\rho _{b\pm }) &=&-\frac{\pi (\rho _{b}^{4}+\ell
^{2}\rho _{b}^{2}+N^{2}(\ell ^{2}+3N^{2}))}{\rho _{b}}\Bigg|\frac{\rho _{b}}{%
3\rho _{b}^{2}-3N^{2}-\ell ^{2}}\Bigg|  \label{TNDS4dIBolt} \\
&=&-\frac{\pi }{216}\left[ \frac{(q^{2}\ell ^{2}+72N^{2})\ell ^{2}}{N^{2}}%
\pm \frac{(q^{2}\ell ^{4}+144N^{4}+48N^{2}\ell ^{2})^{(3/2)}}{N^{2}q\ell ^{2}%
}\right]  \notag
\end{eqnarray}%
and from (\ref{TNDS4dStot}), the bolt entropy is 
\begin{eqnarray}
S_{C,bolt4d}(\rho _{b}=\rho _{b\pm }) &=&\frac{\pi (3\rho _{b}^{4}-(\ell
^{2}+12N^{2})\rho _{b}^{2}-N^{2}(\ell ^{2}+3N^{2}))}{\rho _{b}}\Bigg|\frac{%
\rho _{b}}{3\rho _{b}^{2}-3N^{2}-\ell ^{2}}\Bigg|  \label{TNDS4dSBolt} \\
&=&\frac{\pi }{72}\left[ \frac{(q^{2}\ell ^{2}+24N^{2})\ell ^{2}}{N^{2}}\pm 
\frac{(q\ell ^{2}-12N^{2})(q\ell ^{2}+12N^{2})\sqrt{q^{2}\ell
^{4}+144N^{4}+48N^{2}\ell ^{2}}}{\ell ^{2}qN^{2}}\right]  \notag
\end{eqnarray}%
It can again be checked that this satisfies the first law, though each
branch must be checked separately. From this entropy, the specific heat can
be found for the bolt; explicitly for each branch, it is given by 
\begin{equation}
C_{C,bolt4d}(\rho _{b\pm })=\frac{\pi }{36N^{2}}\left[ q^{2}\ell ^{4}\pm 
\frac{(144q^{2}\ell ^{4}N^{4}+41472N^{8}+10368N^{6}\ell ^{2}+24N^{2}\ell
^{6}q^{2}+q^{4}\ell ^{8})}{q\ell ^{2}\sqrt{q^{2}\ell
^{4}+144N^{4}+48N^{2}\ell ^{2}}}\right]  \label{TNDS4dCBpm}
\end{equation}

\begin{figure}[tbp]
\centering        
\begin{minipage}[c]{.45\textwidth}
         \centering
         \includegraphics[width=\textwidth]{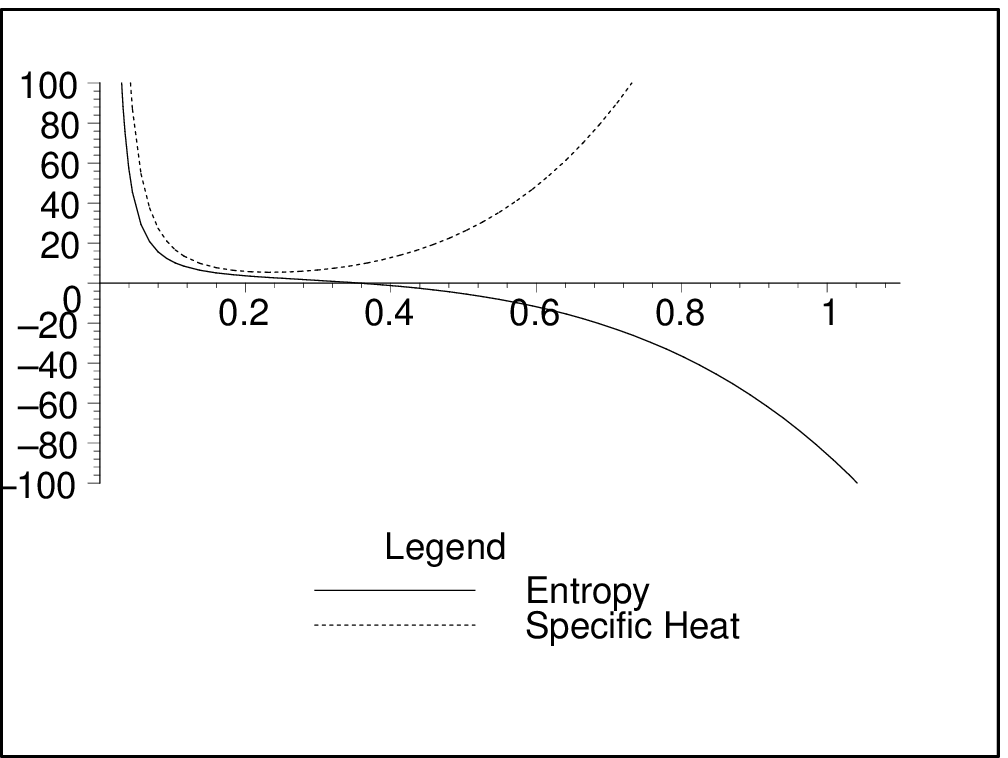}
         \caption{Plot of the upper branch bolt entropy and specific heat (for $q=1$).}
         \label{PlotSCBoltp4d}
\end{minipage}
\begin{minipage}[c]{0.05\textwidth}
\end{minipage}
\begin{minipage}[c]{.45\textwidth}
         \centering
         \includegraphics[width=\textwidth]{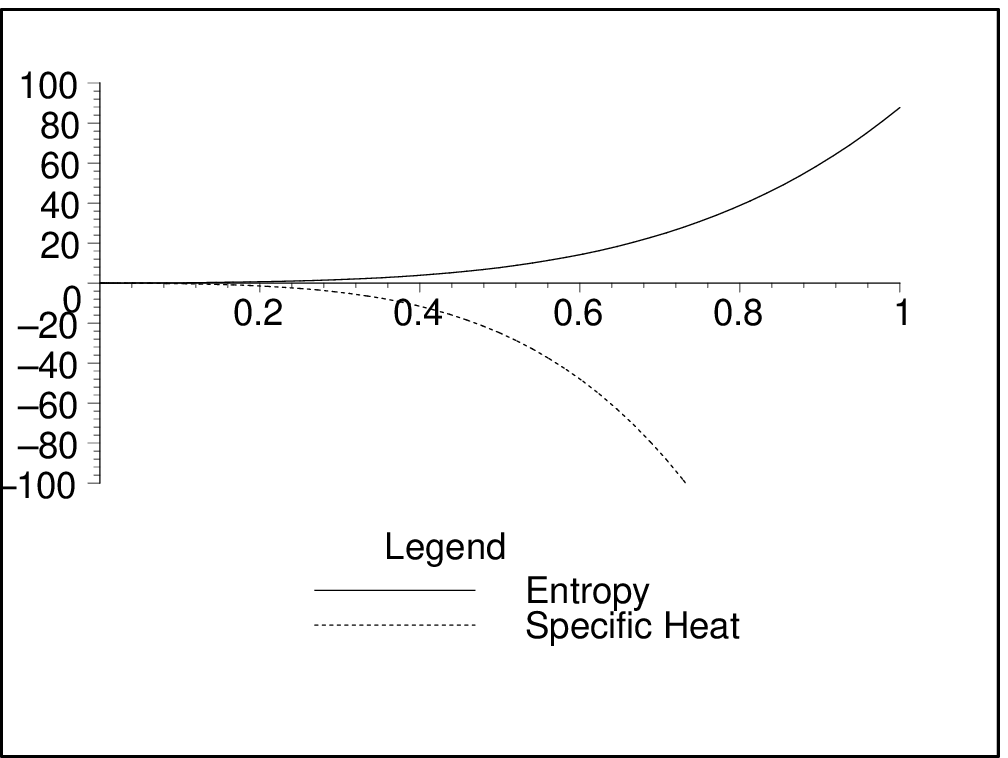}
         \caption{Plot of the lower branch bolt entropy and specific heat (for $q=1$).}
         \label{PlotSCBoltm4d}
\end{minipage}
\end{figure}
Plots of the entropy and specific heat for the upper and lower branch
solutions (for $q=1$) appear in Figures \ref{PlotSCBoltp4d}, \ref%
{PlotSCBoltm4d}. From Figure \ref{PlotSCBoltp4d}, we can see that the
entropy for the upper branch solution is positive for $N<.3562261982\ell $ ,
and the specific heat is always positive; thus the upper branch solution is
thermodynamically stable for $N<.3562261982\ell $. However, for the lower
branch solutions, while the entropy is always positive, the specific heat is
always negative, and so the lower branch bolt solution is always
thermodynamically unstable. Note that this trend continues for $q>1$. Note
also that the lower branch entropy is greater than the pure dS value for $%
N>.3716679966\ell $, showing the lower branch violates the N-bound above
this value of $N$. Similarly, note that the upper branch entropy violates
the N-bound for $N<.2180098653\ell $.

\section{Six Dimensional Analysis}

\label{sec:6d}

\subsection{R-approach in 6 dimensions}

The ($5+1$) dimensional form of the metric ((\ref{TBDSgen}), with $k=2$),
will have in the R-approach 
\begin{equation}
V(\tau )=\frac{3\tau ^{6}+(-\ell ^{2}+15n^{2})\tau ^{4}+3n^{2}(-2\ell
^{2}+15n^{2})\tau ^{2}-3n^{4}(-\ell ^{2}+5n^{2})+6m\tau \ell ^{2}}{3(\tau
^{2}+n^{2})^{2}\ell ^{2}}  \label{FtBolt6}
\end{equation}%
where $n$ is the non-vanishing NUT charge and $\Lambda ={\frac{10}{\ell ^{2}}%
}$. The coordinate $t$ parameterizes an $S^{1}$ Hopf fibered over the
non-vanishing $S^{2}\times S^{2}$ base space, parameterized by $(\theta
_{1},\phi _{1},\theta _{2},\phi _{2})$ . \ It must have periodicity$\frac{%
12\pi \left| n\right| }{q}$ to avoid conical singularities, where $k$ is a
positive integer. The geometry of a constant-$\tau $ surface is that of a
Hopf fibration of $S^{1}$ over $S^{2}\times S^{2}$ which is a well defined
hypersurface in spacetime where $V(\tau )>0$ outside of the past/future
cosmological horizons. The spacelike Killing vector $\partial /\partial t$\
has a fixed point set where $V(\tau _{c})=0$\ whose topology is that of a $%
S^{2}\times S^{2}$ base space. Since $\frac{\partial }{\partial \phi _{1}}$
and $\frac{\partial }{\partial \phi _{2}}$\ are Killing vectors, for any
constant $(\phi _{1},\phi _{2})$-slice near the horizon $\tau =\tau _{c}$\
additional conical singularities will be introduced in the $(t,\tau )$\
Euclidean section unless $t$\ has period 
\begin{equation}
\beta _{R,6d}=\frac{4\pi }{\left| V^{\prime }(\tau _{c})\right| }
\label{betatb6}
\end{equation}%
This period must be equal to $\frac{12\pi \left| n\right| }{q}$, which
forces $\tau _{c}=\tau _{c}^{\pm }$\ where 
\begin{equation}
\tau _{c}^{\pm }=\frac{q\ell ^{2}\pm \sqrt{q^{2}\ell
^{4}-900n^{4}+180n^{2}\ell ^{2}}}{30n}  \label{taus}
\end{equation}%
and the spacetime exists only for the following NUT charges: 
\begin{equation}
\left| n\right| \leq \ell \frac{\sqrt{90+30\sqrt{q^{2}+9}}}{30}  \label{nut6}
\end{equation}%
\bigskip The mass parameters are given by 
\begin{equation}
m_{R}=-\frac{3\tau _{c}^{6}-\tau _{c}^{4}(\ell ^{2}-15n^{2})-\tau
_{c}^{2}n^{2}(6\ell ^{2}-45n^{2})+3n^{4}(\ell ^{2}-5n^{2})}{6\ell ^{2}\tau
_{c}}  \label{masesTB6}
\end{equation}

The conserved mass and the action near future infinity are found to be 
\begin{equation}
\mathfrak{M}_{R,6d}=-8\pi m_{R}-\frac{\pi }{54\ell ^{2}\tau } (2205n^{4}\ell
^{2}-10773n^{6}-\ell ^{6}-63n^{2}\ell ^{4})+\mathcal{O}\left( \frac{1}{\tau
^{2}}\right)  \label{TBmass6}
\end{equation}

\begin{equation}
I_{R,6d}=-\frac{2\beta_{R,6d}\pi }{3\ell^{2}}(3\tau_{c}^{5}+10n^{2}
\tau_{c}^{3} + 15n^{4}\tau_{c} + 3m_{R}\ell^{2}) + \mathcal{O}\left( \frac{1%
}{ \tau }\right)  \label{TBaction6}
\end{equation}

Applying the Gibbs-Duhem relation $S_{R,6d}=\beta _{R,6d}\mathfrak{M}%
_{R,6d}-I_{R,6d}$, the total entropy at future infinity can be found 
\begin{equation}
S_{R,6d}=\frac{2\pi \beta _{R,6d}(3\tau _{c}^{5}+10n^{2}\tau
_{c}^{3}+15n^{4}\tau _{c}-9m_{R}\ell ^{2})}{3\ell ^{2}}  \label{TBentropy6}
\end{equation}%
where $\beta _{R,6d}$ is given by: 
\begin{equation}
\beta _{R,6d}=\frac{6\pi (\tau _{c}^{2}+n^{2})^{3}\ell ^{2}}{\left| 3\tau
_{c}^{7}+9\tau _{c}^{5}n^{2}+\tau _{c}^{3}n^{2}(4\ell ^{2}-15n^{2})-9m\ell
^{2}\tau _{c}^{2}+n^{4}\tau _{c}(75n^{2}-12\ell ^{2})+3m\ell
^{2}n^{2}\right| }  \label{beta6}
\end{equation}

\begin{figure}[tbp]
\centering       
\begin{minipage}[c]{.40\textwidth}
         \centering
         \includegraphics[width=\textwidth]{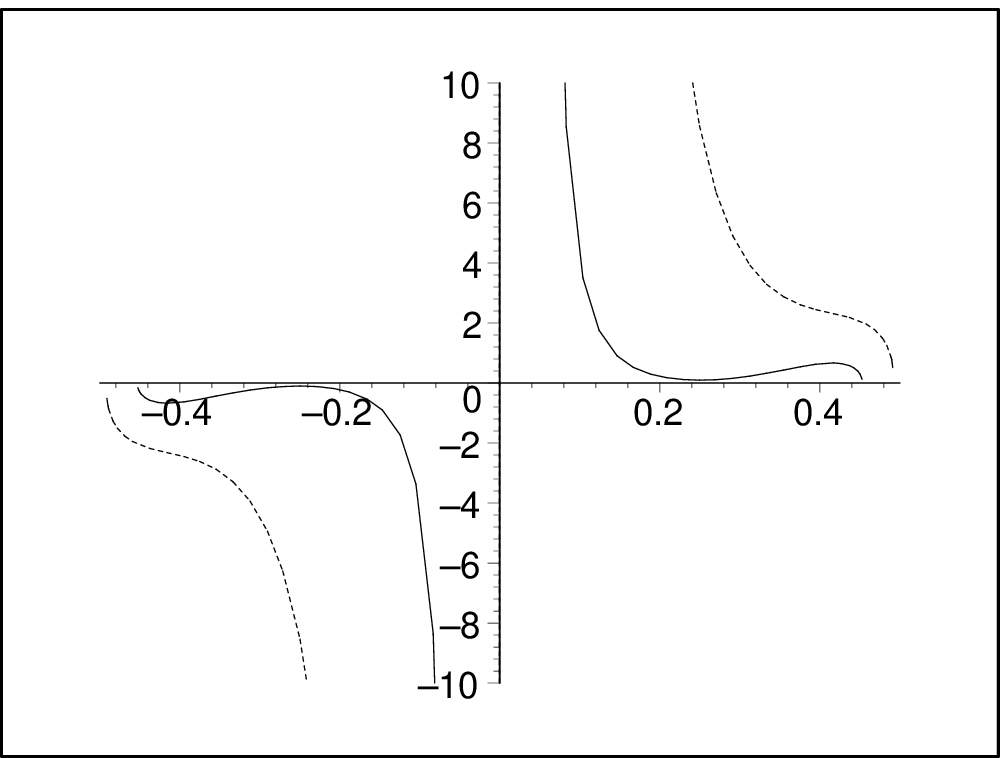}
         \caption{Mass of R$_{6}^{+}$ with $q=1$ (solid) and $q=3$ (dotted).}
         \label{mass6pos}
\end{minipage}\begin{minipage}[c]{0.05\textwidth}
\end{minipage}%
\begin{minipage}[c]{.40\textwidth}
         \centering
         \includegraphics[width=\textwidth]{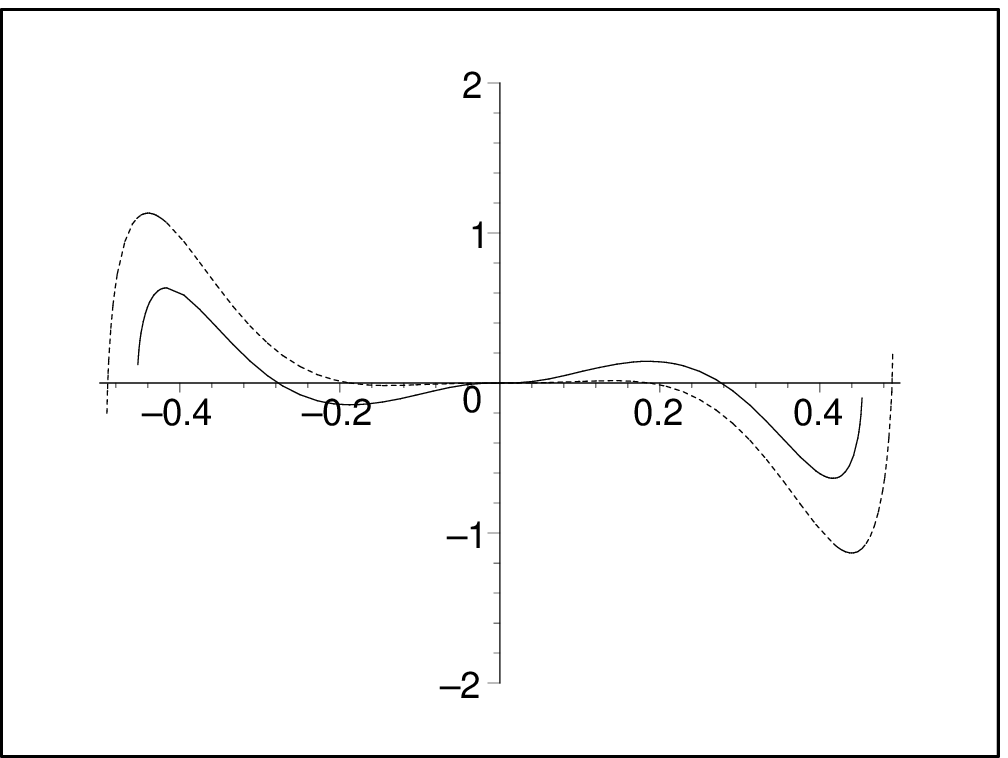}
         \caption{Mass of R$_{6}^{-}$ with $q=1$ (solid) and $q=3$ (dotted).}
         \label{mass6neg}
    \end{minipage}
\end{figure}

\begin{figure}[tbp]
\centering       
\begin{minipage}[c]{.40\textwidth}
         \centering
         \includegraphics[width=\textwidth]{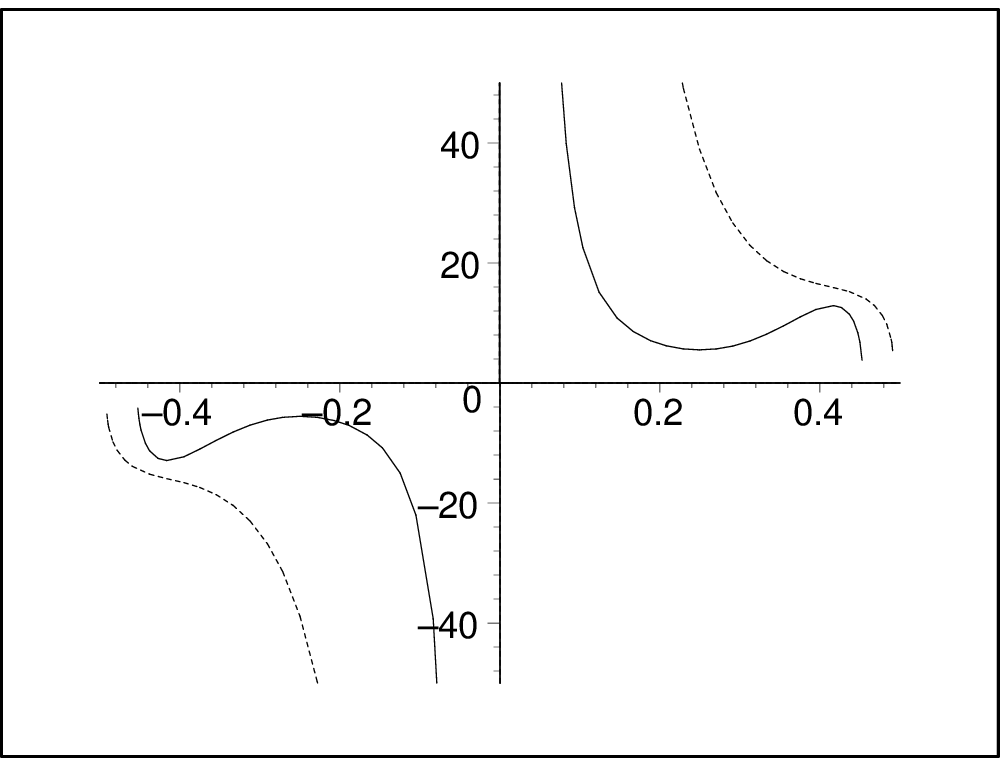}
         \caption{Entropy of R$_{6}^{+}$ with $q=1$ (solid) and $q=3$ (dotted).}
         \label{ent6pos}
\end{minipage}\begin{minipage}[c]{0.05\textwidth}
    \end{minipage}%
\begin{minipage}[c]{.40\textwidth}
         \centering
         \includegraphics[width=\textwidth]{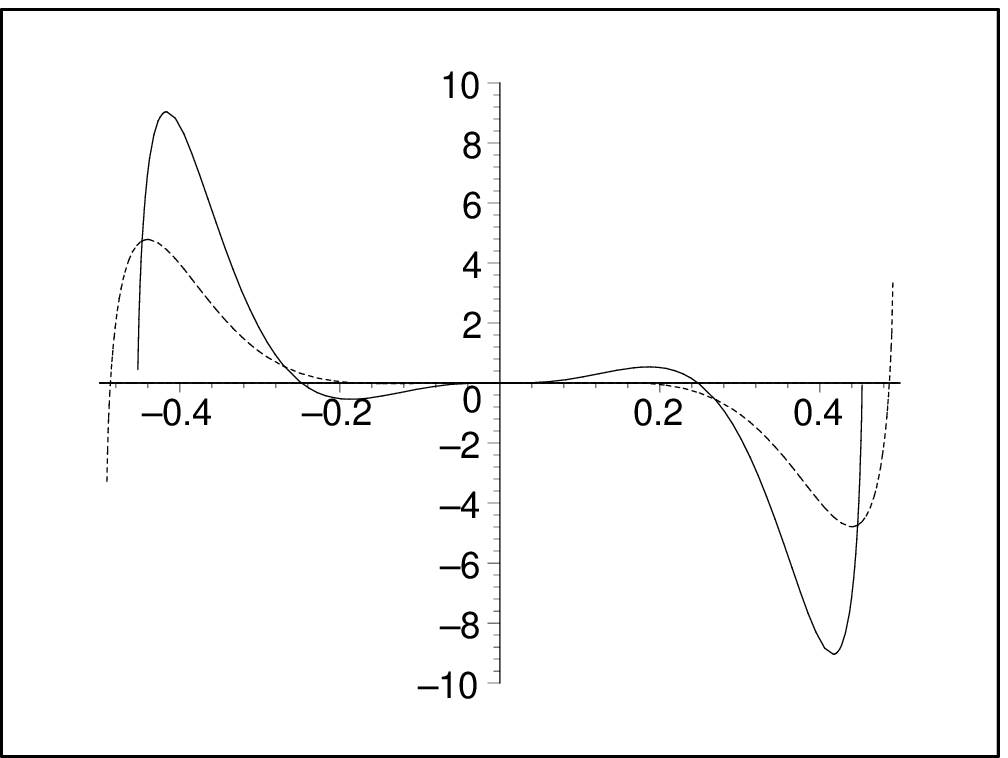}
         \caption{Entropy of R$_{6}^{-}$ with $q=1$ (solid) and $q=3$ (dotted).}
         \label{ent6neg}
    \end{minipage}
\end{figure}

Figures (\ref{mass6pos}), (\ref{mass6neg}), (\ref{ent6pos}) and (\ref%
{ent6neg}) show the conserved masses and entropies for two different
branches of six dimensional R-approach spacetime with $q=1$ and $q=3$.


Figure (\ref{mass6pos}) shows that for all positive NUT charge, R$_{6}^{+}$
has a positive mass. The mass of R$_{6}^{-}$ (with $q=1$) for $%
n<0.27731503405\ell $ is also positive.

Using equations (\ref{TBmass6}), (\ref{masesTB6}) and (\ref{taus}), the R
mass is 
\begin{eqnarray}
\mathfrak{M}_{R,6d}^{\pm }(\tau _{0}=\tau _{0}^{\pm }) &=&\frac{\pm 2\pi }{%
759375n^{5}\ell ^{2}}\{\sqrt{q^{2}\ell ^{4}+180n^{2}\ell ^{2}-900n^{4}}%
(810000n^{8}-54000n^{6}\ell ^{2}  \label{TBDS6mass} \\
&&-1350n^{4}\ell ^{4}+450n^{4}q^{2}\ell ^{4}+60n^{2}q^{2}\ell ^{6}+q^{4}\ell
^{8})\pm 150n^{2}q^{3}\ell ^{8}\pm q^{5}\ell ^{10}\}  \notag
\end{eqnarray}%
and from (\ref{TBentropy6}), the entropy is 
\begin{eqnarray}
S_{R,6d}^{\pm } &=&\frac{\pm \pi ^{2}(\pm q^{2}\ell ^{2}\pm 90n^{2}+q\sqrt{%
q^{2}\ell ^{4}+180n^{2}\ell ^{2}-900n^{4}})^{3}}{101250n^{4}q\ell ^{2}}%
\{(q^{4}\ell ^{8}  \notag \\
&&+90n^{2}\ell ^{6}q^{2}+300n^{4}q^{2}\ell ^{4}-27000n^{6}\ell
^{2}+540000n^{8})  \label{TBDS6entropy} \\
&&\sqrt{q^{2}\ell ^{4}+180n^{2}\ell ^{2}-900n^{4}}\mp 150n^{4}q^{3}\ell
^{6}\pm 4050n^{4}\ell ^{6}q\pm q^{5}\ell ^{10}\pm 180n^{2}\ell ^{8}q^{3}\}/ 
\notag \\
&&\{\pm (60750n^{6}q^{2}-270n^{2}q^{4}\ell ^{4}+675n^{4}q^{4}\ell
^{2}-q^{6}\ell ^{6}-182250n^{6})-  \notag \\
&&\sqrt{q^{2}\ell ^{4}+180n^{2}\ell ^{2}-900n^{4}}(q^{5}\ell
^{4}+6075n^{4}q+180n^{2}\ell ^{2}q^{3}-225n^{4}q^{3})\}  \notag
\end{eqnarray}%
The entropy for both branches satisfies the first law $dS_{R,6d}^{\pm
}=\beta _{R,6d}^{\pm }d\mathfrak{M}_{R,6d}^{\pm }$.

\begin{figure}[tbp]
\centering        
\begin{minipage}[c]{.40\textwidth}
        \centering
        \includegraphics[width=\textwidth]{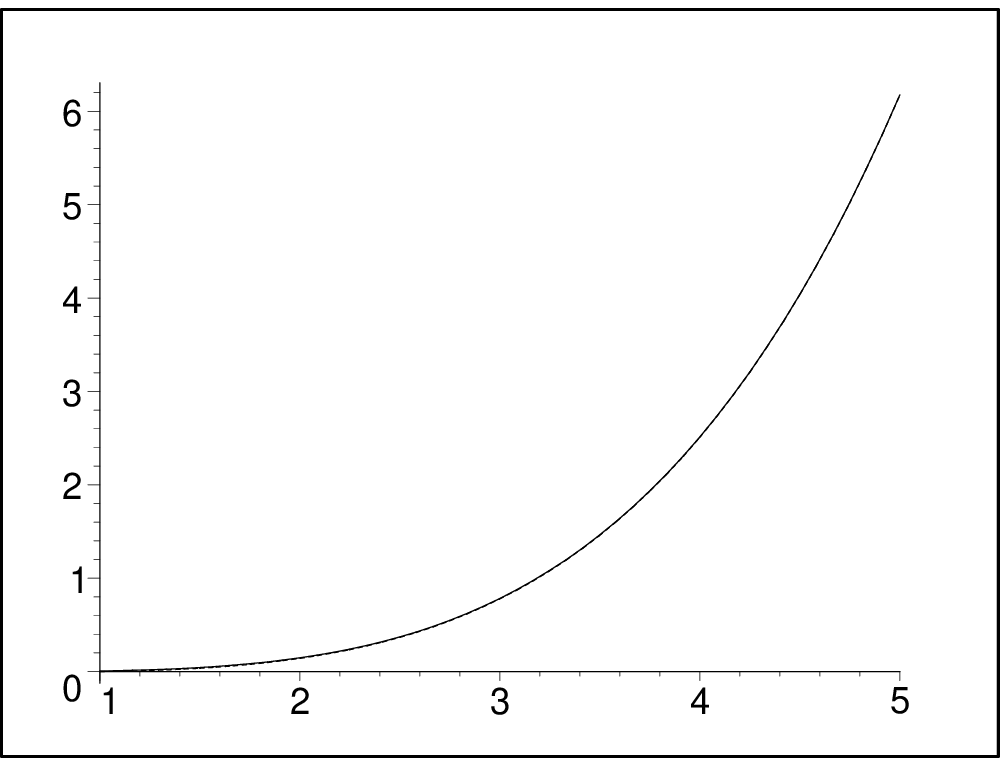}
        \caption{$c$-function of R$_{6}^{+}$  solution
        versus $\tau$ with different values of $q=1$ (solid) and $q=3$ (dotted). The two plots overlap.}
         \label{C6pos}
\end{minipage}
\begin{minipage}[c]{0.05\textwidth}
\end{minipage}
\begin{minipage}[c]{.40\textwidth}
        \centering
        \includegraphics[width=\textwidth]{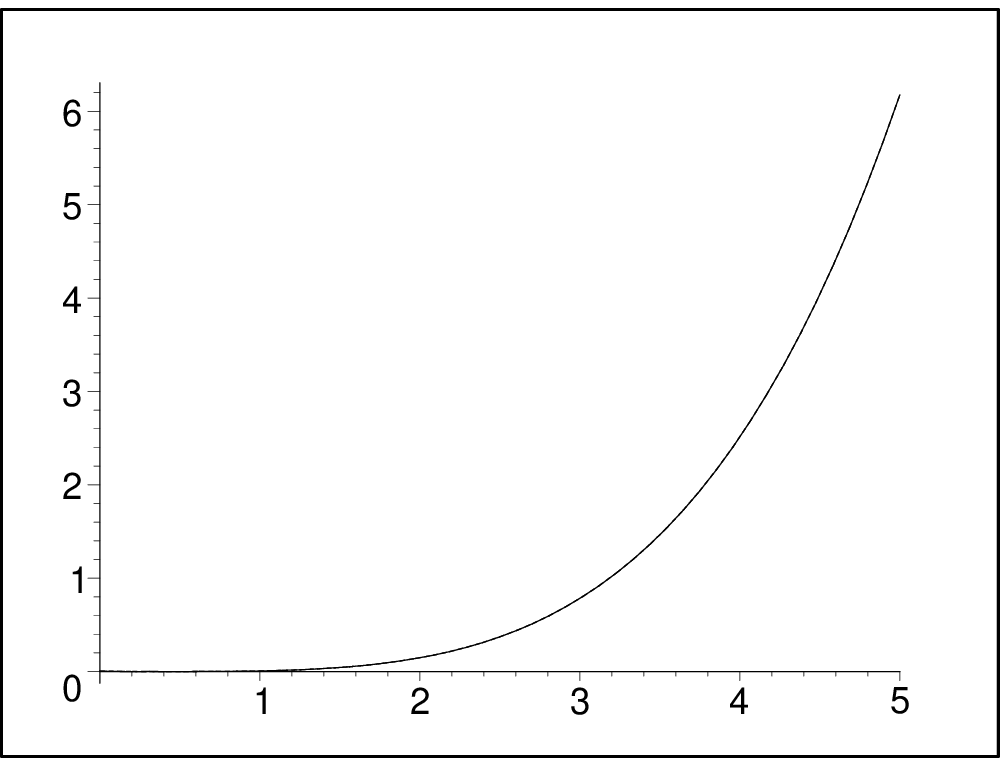}
        \caption{$c$-function of R$_{6}^{-}$  solution
        versus $\tau$ with different values of $q=1$ (solid) and $q=3$ (dotted). The two plots overlap.}
        \label{C6neg}
    \end{minipage}
\end{figure}

The six-dimensional $c$-function is given by 
\begin{equation}
c=\left( G_{\mu \nu }n^{\mu }n^{\nu }\right) ^{-2}=\frac{1}{(G_{\tau \tau
})^{2}}  \label{cfunction6d}
\end{equation}%
where $n^{\mu }$\ is the unit normal vector to a constant $\tau -$slice. In
figures (\ref{C6pos}) and (\ref{C6neg}), the diagrams of a Taub-Bolt-dS
spacetimes $c$-functions outside the cosmological horizon with $\ell =1$ and 
$n=0.25$ for two cases $q=1$ and $3$ are plotted. 

As one can see from these figures, outside the cosmological horizon, the $c$
-function is a monotonically increasing function of coordinate $\tau ,$
showing the expansion of a constant $\tau $-surface of the metric (\ref%
{TBDSgen}) outside of the cosmological horizon$.$ We note that the behavior
of the $c$-function is rather insensitive to $q.$

\subsection{C-approach in 6 dimensions}

In this approach the metric now has the form (\ref{TNDSgen}) with $k=2,d=5$,
and will have 
\begin{equation}
F(\rho )=\frac{3\rho ^{6}-(\ell ^{2}+15N^{2})\rho ^{4}+3N^{2}(2\ell
^{2}+15N^{2})\rho ^{2}+3N^{4}(\ell ^{2}+5N^{2})+6m\rho \ell ^{2}}{3(\rho
^{2}-N^{2})^{2}\ell ^{2}}  \label{TNDS6dFr}
\end{equation}%
where $N$ is again the non-vanishing NUT charge and the cosmological
constant is now given by $\Lambda ={\textstyle\frac{10}{\ell ^{2}}}$. The
periodicity condition becomes 
\begin{equation}
\beta_{C,6d}=\frac{4\pi }{\left| F^{\prime }(\rho )\right| }=\frac{12\pi |N| 
}{q}  \label{TNDS6dbeta}
\end{equation}%
to avoid conical singularities in 6 dimensions.

The geometric interpretation is fraught with the same difficulties as its 4
dimensional counterpart. \ Notwithstanding these issues, we shall proceed as
before.

The action follows from (\ref{TNDSactiongeneral}) with $k=2$ 
\begin{equation}
I_{C,6d}=-\frac{2\pi \beta (3\rho _{+}^{5}+15N^{4}\rho _{+}+3m\ell
^{2}-10N^{2}\rho ^{3})}{3\ell ^{2}}  \label{TNDS6dItot}
\end{equation}%
with $\rho _{+}$ the largest positive root of $F(\rho )$, determined by the
fixed point set of $\partial _{T}$, and $m=m_{C,6d}$ the mass parameter for
six dimensions.

Working at future infinity, the conserved mass is 
\begin{equation}
\mathfrak{M}_{C,6d}=-8\pi m_{C,6d}-\frac{\pi (63\ell ^{4}N^{2}+2205N^{4}\ell
^{2}+10773N^{6}-\ell ^{6})}{54\rho \ell ^{2}}+\mathcal{O}\left( \frac{1}{%
\rho ^{2}}\right)  \label{TNDS6dMass}
\end{equation}%
and the total entropy is 
\begin{equation}
S_{C,6d}=\frac{2\pi \beta (3\rho _{+}^{5}-10N^{2}\rho _{+}^{3}+15N^{4}\rho
_{+}-9m\ell ^{2})}{3\ell ^{2}}  \label{TNDS6dStot}
\end{equation}

These equations are generic, and can be analyzed for the specific
6-dimensional Taub-NUT-C and Taub-Bolt-C cases. When $\rho_{+}=N$, $F(\rho
=N)=0$ and the fixed point set of $\partial _{T}$ is 2-dimensional, giving
the NUT solution; when $\rho _{+}=\rho _{b\pm }>N$, the fixed point set is
4-dimensional, giving the bolt solutions.

\subsubsection{Taub-NUT-C Solution}

For the NUT solution, $\rho _{+}=N$, and the NUT mass is 
\begin{equation}
m_{C,n6}=-\frac{4N^{3}(\ell ^{2}+6N^{2})}{\ell ^{2}}  \label{TNDS6dmn}
\end{equation}%
It is easily seen from this that $m_{C,n6}$ is always negative, which (from (%
\ref{TNDS6dMass})) will give a positive conserved mass at future infinity.
In the flat space limit, the NUT mass will go to $-{\ \textstyle\frac{4}{3}}%
N^{3}$, and in the high temperature limit, $m_{C,n6}$ goes to 0.

The period in six dimensions ($q=1$) is $\beta = 12\pi N$, so from (\ref%
{TNDS6dItot},\ref{TNDS6dStot}), 
\begin{eqnarray}
I_{C,NUT6d} & = & \frac{ 32 \pi^2 N^4 (\ell^2 + 4 N^2)}{\ell^2}
\label{TNDS6dINUT} \\
S_{C,NUT6d} & = & \frac{ 32 \pi^2 N^4 (3\ell^2 + 20 N^2)}{\ell^2}
\label{TNDS6dSNUT}
\end{eqnarray}
(\ref{TNDS6dSNUT}) and (\ref{TNDS6dMass}) with $m=m_{C,n6}$ can be shown to
satisfy the first law $dS = \beta d\mathfrak{M}$. In the flat space limit, $
I_{C,NUT6d} \rightarrow 32 \pi^2 N^4$, and $S_{C,NUT6d} \rightarrow 96 \pi^2
N^4$. Both the action and the entropy go to 0 in the high temperature limit.

The specific heat in six dimensions can also be calculated, 
\begin{equation}
C_{C,NUT6d} = -\frac{ 384 \pi^2 N^4 (\ell^2 + 10 N^2)}{\ell^2}
\label{TNDS6dCNUT}
\end{equation}
where this will go to $-384 \pi^2 N^4$ in the flat space limit, and will go
to 0 in the high temperature limit.

\begin{figure}[tbp]
\centering       
\begin{minipage}[c]{.45\textwidth}
         \centering
         \includegraphics[width=\textwidth]{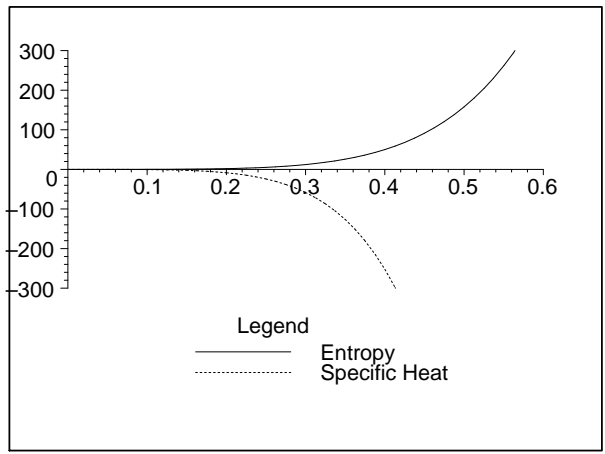}
         \caption{Plot of the NUT entropy and specific heat vs. $N$ for (5+1) dimensions.}
         \label{PlotSCNUT6d}
     \end{minipage}
\end{figure}

In six dimensions, it can be seen (see Figure \ref{PlotSCNUT6d}) that the
entropy is always positive, and the specific heat is always negative. This
is opposite to what occurs in four dimensions, though as in four dimensions,
this means that the NUT solution is thermodynamically unstable.

\subsubsection{Taub-Bolt-C Solution}

In this case the fixed point set of $\partial _{T}$ is 4 dimensional, giving 
$\rho _{+}=\rho _{b\pm }>N$. The conditions for a regular bolt solution are
now (i) $F(\rho )=0$, and (ii) $F^{\prime }(\rho )=\pm {\textstyle\frac{q}{3N%
}}$ (where (ii) comes from the second equality in (\ref{TNDS6dbeta})). From
(i), the bolt mass is given by 
\begin{equation}
m_{C,b6}=-\frac{3\rho _{b}^{6}-(\ell ^{2}+15N^{2})\rho _{b}^{4}+N^{2}(6\ell
^{2}+45N^{2})\rho _{b}^{2}+3N^{4}(\ell ^{2}+5N^{2})}{6\ell ^{2}\rho _{b}}
\label{TNDS6dmb}
\end{equation}%
$\rho _{b\pm }$ is given by (ii$^{+}$) 
\begin{equation}
\rho _{b\pm }=\frac{q\ell ^{2}\pm \sqrt{q^{2}\ell ^{4}+900N^{4}+180N^{2}\ell
^{2}}}{30N}  \label{TNDS6dtbpm}
\end{equation}%
Note again that the discriminant of $\rho _{b\pm }$ will always be positive,
and so there will be no limit on $N$ (except $N>0$). The flat space and high
temperature limits of $\rho _{b+}$ are infinite; the flat space limit of $%
\rho _{b-}$ is $-{\textstyle\frac{3N}{q}}$, and the high temperature limit
is $0$.

The period for the bolt is found from the first equality in (\ref{TNDS6dbeta}%
) 
\begin{equation}
\beta_{C,Bolt6d} = 6 \pi \left| \frac{ (\rho_b^2-N^2)^3 \ell^2 }{ 3 \rho_b^7
- 9 \rho_b^5 N^2 - N^2 ( 4 \ell^2 + 15 N^2 ) \rho_b^3 - 9 m \ell^2 \rho_b^2
- N^4( 12 \ell^2 + 75 N^2) \rho_b - 3 m \ell^2 N^2} \right|
\label{TNDS6dbet}
\end{equation}
The temperature of the NUT and bolt solutions can again be shown to be the
same, by substituting in $m=m_{C,b6}$ and either of $\rho_{b\pm}$.

\begin{figure}[tbp]
\centering        
\begin{minipage}[c]{.45\textwidth}
         \centering
         \includegraphics[width=\textwidth]{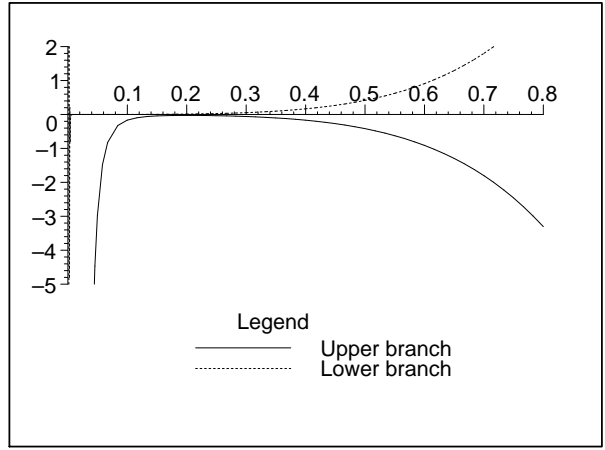}
    \end{minipage}
\caption{Plot of the upper ($\protect\rho _{b}=\protect\rho _{b+}$) and
lower ( $\protect\rho _{b}=\protect\rho _{b-}$) bolt masses $m_{b\pm }$ ($%
q=1 $) for six dimensions.}
\label{Plotmbpm6d}
\end{figure}

Substituting $\rho _{b}=\rho _{b\pm }$ into (\ref{TNDS6dmb}), we can plot
the mass vs. $N$ (see Figure \ref{Plotmbpm6d}). The upper branch mass is
always negative, and the lower branch mass is always positive. Since the
conserved mass is again negative at future infinity (\ref{TNDS6dMass}), this
will mean that in six dimensions, the bolt upper branch conserved mass will
always be positive, and the bolt lower branch conserved mass negative. Note
that this is different than the four dimensional case, where the upper
branch solution varied from positive to negative.

Now, from (\ref{TNDS6dItot}), using (\ref{TNDS6dmb}) and (\ref{TNDS6dtbpm}),
the action is 
\begin{eqnarray}
I_{C,Bolt6d} &=&-\frac{4\pi ^{2}(3\rho _{b}^{6}+(\ell ^{2}-5N^{2})\rho
_{b}^{4}-N^{2}(6\ell ^{2}+15N^{2})\rho _{b}^{2}-3N^{4}(\ell ^{2}+5N^{2}))}{%
3|5N^{2}-5\rho _{b}^{2}+\ell ^{2}|}  \label{TNDS6dIbpm} \\
&=&\frac{2\pi ^{2}}{253125}\left[ -\frac{\ell
^{4}(20250N^{4}+750N^{4}q^{2}+300q^{2}\ell ^{2}N^{2}+q^{4}\ell ^{4})}{N^{4}}%
\right.  \notag \\
&&\left. \pm \frac{(-q^{2}\ell ^{4}+600N^{4}-30N^{2}\ell ^{2})(q^{2}\ell
^{4}+900N^{4}+180N^{2}\ell ^{2})^{3/2}}{\ell ^{2}qN^{4}}\right]  \notag
\end{eqnarray}%
and from (\ref{TNDS6dStot}), the entropy is 
\begin{eqnarray}
S_{C,Bolt6d} &=&\frac{4\pi ^{2}(15\rho _{b}^{6}-(3\ell ^{2}+65N^{2})\rho
_{b}^{4}+3N^{2}(6\ell ^{2}+55N^{2})\rho _{b}^{2}+9N^{4}(\ell ^{2}+5N^{2}))}{%
3|5N^{2}-5\rho _{b}^{2}+\ell ^{2}|}  \label{TNDS6dSbpm} \\
&=&\frac{2\pi ^{2}}{50625}\left[ \frac{(q^{4}\ell ^{4}+180N^{2}q^{2}\ell
^{2}+150N^{4}q^{2}+4050N^{4})\ell ^{4}}{N^{4}}\right.  \notag \\
&&\left. \pm \frac{(q^{4}\ell ^{8}+90N^{2}q^{2}\ell ^{6}-300N^{4}q^{2}\ell
^{4}+27000N^{6}\ell ^{2}+540000N^{8})\sqrt{q^{2}\ell
^{4}+900N^{4}+180N^{2}\ell ^{2}}}{N^{4}q\ell ^{2}}\right]  \notag
\end{eqnarray}%
This entropy does satisfy the first law, though note that both branches must
be checked separately. The specific heat (explicitly for each branch) in six
dimensions is given by 
\begin{eqnarray}
C_{C,Bolt6d} &=&\frac{8\pi ^{2}}{50625}\left[ \frac{\ell
^{6}q^{2}(90N^{2}+q^{2}\ell ^{2})}{N^{4}}\right.  \notag \\
&&\left. \pm \frac{(q\ell ^{2}+30N^{2})(q\ell ^{2}-30N^{2})}{N^{4}q\ell ^{2}%
\sqrt{q^{2}\ell ^{4}+900N^{4}+180N^{2}\ell ^{2}}}\Big(q^{4}\ell
^{8}+180N^{2}q^{2}\ell ^{6}+1350N^{4}q^{2}\ell ^{4}\right.  \notag \\
&&\left. +4050N^{4}\ell ^{4}+162000N^{6}\ell ^{2}+810000N^{8}\Big)\right]
\label{TNDS6dcbpm}
\end{eqnarray}%
\begin{figure}[tbp]
\centering       
\begin{minipage}[c]{.4\textwidth}
         \centering
         \includegraphics[width=\textwidth]{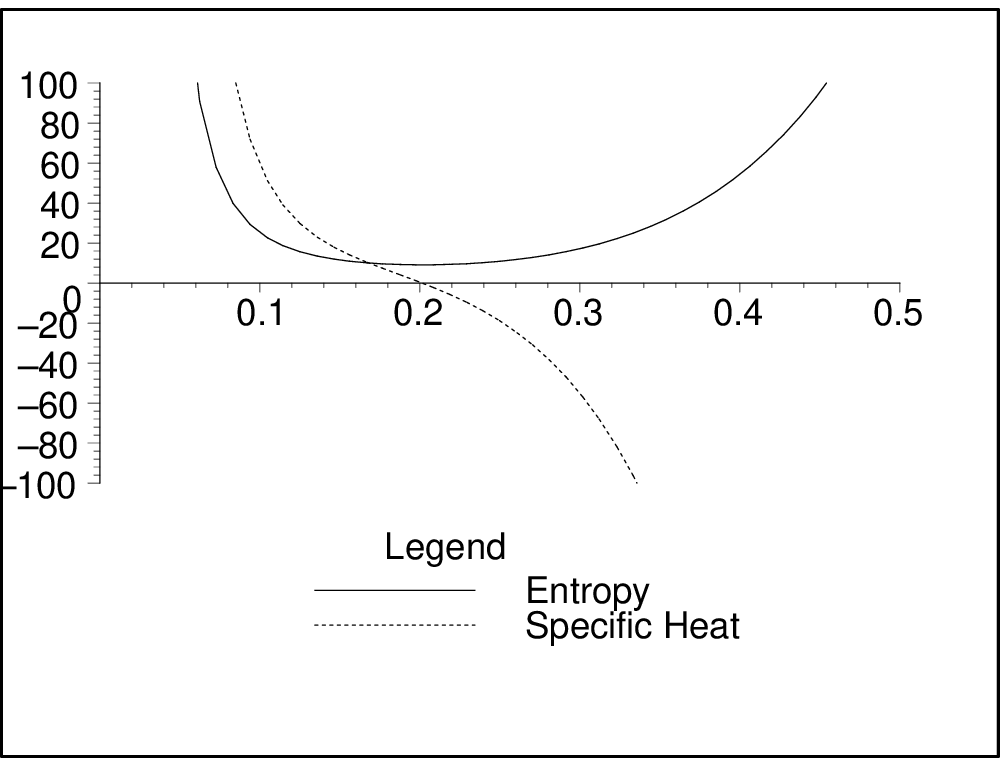}
         \caption{Plot of the upper branch bolt entropy and specific heat
         (for $q=1$) for six dimensions.} \label{PlotSCBoltp6d}
\end{minipage}\begin{minipage}[c]{0.05\textwidth}
\end{minipage}%
\begin{minipage}[c]{.4\textwidth}
         \centering
         \includegraphics[width=\textwidth]{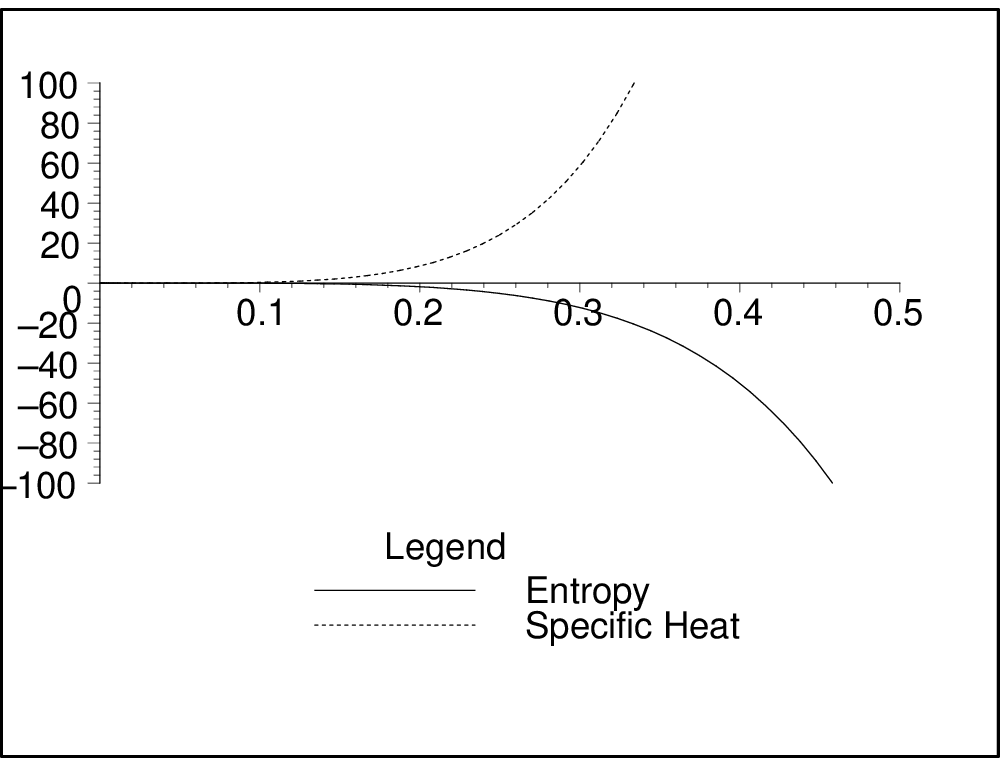}
         \caption{Plot of the lower branch bolt entropy and specific heat (for $q=1$) for six dimensions.}
         \label{PlotSCBoltm6d}
\end{minipage}
\end{figure}
Plots of the entropy and specific heat for the upper and lower branch
solutions (for $q=1$) appear in Figures \ref{PlotSCBoltp6d}, \ref%
{PlotSCBoltm6d}. From Figure \ref{PlotSCBoltp6d}, we can see that the
entropy for the upper branch solution is always positive, and the specific
heat is positive for $N<.2014312523~\ell $; thus the upper branch solution
is thermodynamically stable for $N$ less than this. However, for the lower
branch solutions, the entropy is always negative and the specific heat
always positive, and so the lower branch bolt solution is always
thermodynamically unstable. Note that this trend continues for $q>1$.

\section{Discussion}

We have extended the use of the path-integral formalism to include quantum
correlations between timelike histories. \ By employing this formalism in
the semiclassical approximation we have been able to extend our notions of
conserved quantities (such as mass and angular momentum), actions and
entropies outside of cosmological horizons. \ Applying this formalism to
Schwarzschild de Sitter spacetimes we find that the values of these
quantities are in accord with our physical expectations, as previously shown
in refs.\cite{bala,cai,GM}.

When we extend this formalism to NUT-charged spacetimes we find that the
situation is considerably modified. First, NUT-charged spacetimes present us
with two possible ways (the R-approach and the C-approach) in which we can
apply our formalism, depending on how the spacetime is analytically
continued. \ \ Moreover, there exist broad ranges of parameter space for
which NUT-charged spacetimes violate both the maximal mass conjecture and
the N-bound, in both four dimensions and in higher dimensions. \ We present
in tables 1 and 2 the results for dimensions $4,6,8,10$ and general $(d+1)$
dimensions. \ We find the thermodynamic behaviour for the $4k$-dimensional
spacetimes to be qualitatively similar in the R-approach, with the lower
branch entropy always negative, and the upper branch solutions always having
a range of $n$ in which both the entropy and the specific heat are positive.
Likewise the $(4k+2)$-dimensional spacetimes have qualitatively similar
thermodynamics, behaving as illustrated in figures (\ref{mass6pos}), (\ref%
{mass6neg}), (\ref{ent6pos}) and (\ref{ent6neg}). \ \ In the C-approach we
likewise find a similarity in the thermodynamics of the $4k$-dimensional
spacetimes, distinct from the common behaviour of the $(4k+2)$-dimensional
ones.

These results suggest that there may be some limitations to the application
of the holographic conjecture to spacetimes with $\Lambda >0$. \ For
example, one implication of the N-bound (and the maximal mass conjecture) is
that a quantum gravity theory with an infinite number of degrees of freedom
(such as M-theory) cannot describe spacetimes with $\Lambda >0$ \cite{bousso}%
. Our results suggest that this obstruction is not necessarily an
obstruction in principle, but can be overcome in spacetimes with NUT charge
that are locally asymptotically dS. \ One might wish to restrict the
appearance of such spacetimes in the spectrum of states of quantum gravity
since they contain contain causality-violating regions with closed timelike
curves. The mechanism for so doing remains an unsolved problem.

The entropy-area relation $S=A/4$\ is satisfied for any black hole in a $%
(d+1)$-dimensional aAdS or aF, where $A$\ is the area of a $(d-1)$%
-dimensional fixed point set of isometry group. However, the entropy can
defined for other kinds of spacetimes in which the isometry group has fixed
points on surfaces of even co-dimension \cite{Haw}. The best examples of
these spacetimes are asymptotically locally flat and asymptotically locally
AdS spacetimes with NUT charge. In these cases when the isometry group has a
two-dimensional fixed set (bolt), the entropy of the spacetime is not given
by the area-entropy relation, since there is a contribution to the entropy,
coming from the Misner string \cite{MannMisner}.

In asymptotically dS spacetimes, the Gibbs-Duhem entropy (\ref{GDoutfinal})
is less than the area of the horizon and respects the N-bound (for the case
of Schwarzschild-dS spacetime, see \cite{GM}). However, for asymptotically
locally dS spacetime with NUT charge, we have an additional contribution to
the entropy (\ref{GDoutfinal}) from the Misner string. Consequently the
entropy need not respect the N-bound, and we find that there are a wide
range of situations where it does not.

\begin{figure}[tbp]
\centering        
\begin{minipage}[c]{.4\textwidth}
         \centering
         \includegraphics[width=\textwidth]{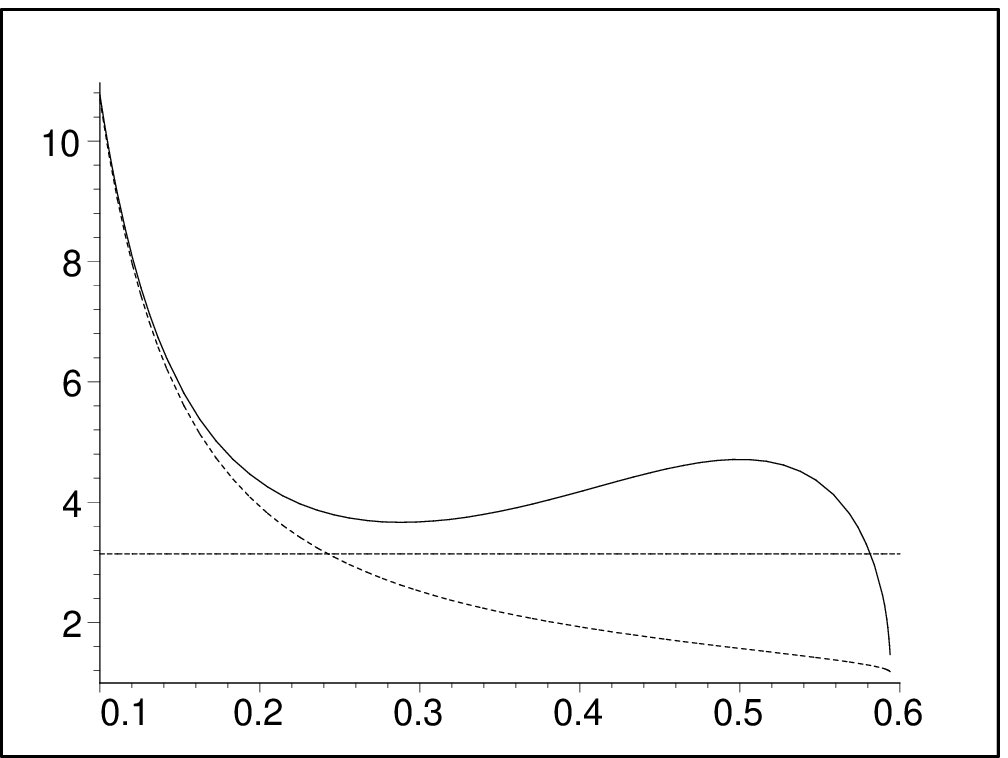}
         \caption{Gibbs-Duhem entropy (solid), cosmological entropy (dotted) and N-bound (dashed)
for positive NUT charge of $R_4^+$}
         \label{Entropiesp}
\end{minipage}
\begin{minipage}[c]{0.05\textwidth}
\end{minipage}
\begin{minipage}[c]{.4\textwidth}
         \centering
         \includegraphics[width=\textwidth]{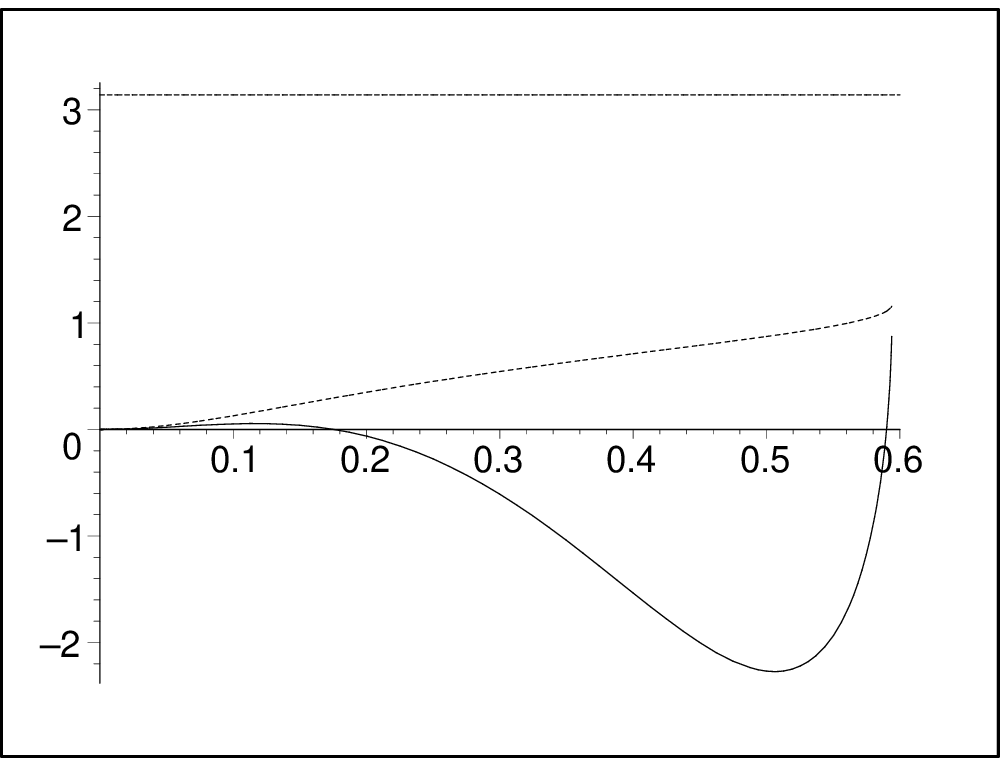}
         \caption{Gibbs-Duhem entropy (solid), cosmological entropy (dotted) and N-bound (dashed)
for positive NUT charge of $R_4^-$}
         \label{Entropiesm}
\end{minipage}
\end{figure}

In fact for positive NUT charge, for R$_{4}^{+},$\ the fixed-$t$\ area of
the cosmological horizon exceeds that of pure dS spacetime for certain range
of values of the NUT charge. Consequently if one interprets the N-bound in
terms of a relationship between horizon areas (as opposed to entropies), we
still find that (within this range) the N-bound is violated. For R$_{4}^{-},$%
\ the fixed-$t$\ area of the cosmological horizon is less than the
cosmological horizon area of pure dS spacetime for all values of NUT charge,
and so the re-interpreted N-bound is respected. In the R$_{4}^{+}$ case, the
Gibbs-Duhem entropy is larger than one-quarter of the horizon area, which in
turn is larger than the cosmological horizon area $\pi \ell ^{2}$ of pure dS
spacetime. In the R$_{4}^{-}$ case, these inequalities are reversed, with $%
\pi \ell ^{2}$ always greater than the Gibbs-Duhem entropy, and both the
Gibbs-Duhem entropy and one-quarter the area of cosmological horizon respect
N-bound. Figures (\ref{Entropiesp}) and (\ref{Entropiesm}) show the
behaviour of entropies for the R$_{4}^{+}$ and R$_{4}^{-}.$%

Both the robustness of our formalism and its physical relevance remain
subjects for future study. \ The entropy defined by the Gibbs-Duhem relation
(\ref{GDoutfinal}) would appear to have the requisite properties: it is
positive and monotonically increasing with conserved mass for the
Schwarzschild de Sitter case, and obeys the first law of thermodynamics for
all cases we have considered so far (indeed, since our definition is built
on the path integral formalism, it is hard to see how it could be
otherwise). \ However the applicability of the second law remains an
outstanding problem: in what sense can we say that the entropy always
increases in any physical process in this context? \ Even more intriguing is
the relationship between this entropy and the underlying degrees of freedom
that it presumably counts. \ 

\bigskip

{\Large Acknowledgments}

This work was supported by the Natural Sciences and Engineering Research
Council of Canada.

\begin{table}[!hbp]
\caption{Summary of General R-approach quantities }
\label{summarytableR}\centering        
\begin{tabular}{|c|c|c|c|c|}
\hline
Dim. & Radial ($\tau_{c\pm}$) & Period & $\mathfrak{M}$ & Action \\ \hline
&  &  &  &  \\ 
4 & $\frac{ \ell^2 \pm \sqrt{\ell^4 - 144n^4 + 48n^2 \ell^2}}{12n} $ & $8\pi
N$ & $-m $ & $-\frac{ \beta (m \ell^2 + \tau_c^3 + 3 \tau_c n^2 ) }{2\ell^2} 
$ \\ 
&  &  &  &  \\ \hline
&  &  &  &  \\ 
6 & $\frac{ \ell^2 \pm \sqrt{\ell^4 - 900n^4 + 180n^2 \ell^2}}{30n} $ & $12
\pi N$ & $-8 \pi m $ & $%
\begin{array}{c}
-\frac{2 \beta \pi }{3 \ell^2} \Big(3 m \ell^2 + 3 \tau_c^5 \\ 
+ 10 n^2 \tau_c^3 + 15 n^4 \tau_c \Big)%
\end{array}
$ \\ 
&  &  &  &  \\ \hline
&  &  &  &  \\ 
8 & $\frac{ \ell^2 \pm \sqrt{\ell^4 - 3136n^4 + 448n^2 \ell^2}}{56n} $ & $16
\pi N $ & $- 48 \pi^2 m $ & $%
\begin{array}{c}
- \frac{ 8 \beta \pi^2 }{5 \ell^2} \Big(5 m \ell^2 + 5 \tau_c^7 \\ 
+ 21 n^2 \tau_c^5 + 35 n^4 \tau_c^3 \\ 
+ 35 n^6 \tau_c \Big)%
\end{array}
$ \\ 
&  &  &  &  \\ \hline
&  &  &  &  \\ 
10 & $\frac{ \ell^2 \pm \sqrt{\ell^4 - 8100n^4 + 900n^2 \ell^2}}{90n} $ & $%
20 \pi N $ & $-256 \pi^3 m $ & $%
\begin{array}{c}
-\frac{ 32 \beta \pi^3 }{35 \ell^2 } \Big( 35 m \ell^2 + 35 \tau_c^9 \\ 
+ 180 n^2 \tau_c^7 + 378 n^4 \tau_c^5 \\ 
+ 420 n^6 \tau_c^3 + 315 n^8 \tau_c \Big)%
\end{array}
$ \\ 
&  &  &  &  \\ \hline
&  &  &  &  \\ 
(d+1) & $\frac{2}{n(d+1)} = \frac{4\pi}{|V^{\prime}(\tau)|} $ & $\frac{
2(d+1) \pi n }{q} $ & $-\frac{ (4 \pi)^k k m }{4\pi} $ & $%
\begin{array}{c}
-\frac{ \beta (4 \pi)^k }{8 \pi \ell^2 } \Big[ m \ell^2 \\ 
+ d \Big( \sum_{i=0}^k \binom{k}{i} \frac{ n^{2i} \tau_c^{2k-2i+1} }{
2k-2i+1 } \Big) \Big]%
\end{array}
$ \\ 
&  &  &  &  \\ \hline
\end{tabular}%
\end{table}

\begin{table}[!hbp]
\caption{Summary of General C-approach quantities }
\label{summarytableC}\centering        
\begin{tabular}{|c|c|c|c|c|}
\hline
Dim. & $\rho_{+} $ & Period & $\mathfrak{M}$ & Action \\ \hline
&  &  &  &  \\ 
4 & $%
\begin{array}{l}
\rho_+ = N \\ 
\rho_{b\pm} = \frac{ \ell^2 \pm \sqrt{\ell^4 + 144N^4 + 48N^2 \ell^2}}{12N}%
\end{array}
$ & $8\pi N$ & $-m $ & $\frac{ \beta (- m \ell^2 - \rho_+^3 + 3 N^2 \rho_+ ) 
}{2 \ell^2} $ \\ 
&  &  &  &  \\ \hline
&  &  &  &  \\ 
6 & $%
\begin{array}{l}
\rho_+ = N \\ 
\rho_{b\pm} = \frac{ \ell^2 \pm \sqrt{\ell^4 + 900N^4 + 180N^2 \ell^2}}{30N}%
\end{array}
$ & $12 \pi N$ & $-8 \pi m $ & $%
\begin{array}{c}
-\frac{2 \beta \pi }{3 \ell^2} \Big(3 m \ell^2 + 3 \rho_+^5 \\ 
- 10 N^2 \rho_+^3 + 15 N^4 \rho_+ \Big)%
\end{array}
$ \\ 
&  &  &  &  \\ \hline
&  &  &  &  \\ 
8 & $%
\begin{array}{l}
\rho_+ = N \\ 
\rho_{b\pm} = \frac{ \ell^2 \pm \sqrt{\ell^4 + 3136N^4 + 448N^2 \ell^2}}{56N}%
\end{array}
$ & $16 \pi N $ & $- 48 \pi^2 m $ & $%
\begin{array}{c}
\frac{ 8 \beta \pi^2 }{5 \ell^2} \Big(-5 m \ell^2 - 5 \rho_+^7 \\ 
+ 21 N^2 \rho_+^5 - 35 N^4 \rho_+^3 \\ 
+ 35 N^6 \rho_+ \Big)%
\end{array}
$ \\ 
&  &  &  &  \\ \hline
&  &  &  &  \\ 
10 & $%
\begin{array}{l}
\rho_+ = N \\ 
\rho_{b\pm} = \frac{ \ell^2 \pm \sqrt{\ell^4 + 8100N^4 + 900N^2 \ell^2}}{90N}%
\end{array}
$ & $20 \pi N $ & $-256 \pi^3 m $ & $%
\begin{array}{c}
-\frac{ 32 \beta \pi^3 }{35 \ell^2 } \Big( 35 m \ell^2 + 35 \rho_+^9 \\ 
- 180 N^2 \rho_+^7 + 378 N^4 \rho_+^5 \\ 
- 420 N^6 \rho_+^3 + 315 N^8 \rho_+ \Big)%
\end{array}
$ \\ 
&  &  &  &  \\ \hline
&  &  &  &  \\ 
(d+1) & $%
\begin{array}{l}
\rho_+ = N \\ 
\frac{2}{N(d+1)} = \frac{4\pi}{|F^{\prime}(\rho)|}%
\end{array}
$ & $\frac{ 2(d+1) \pi n }{q} $ & $-\frac{ (4 \pi)^k k m }{4\pi} $ & $%
\begin{array}{c}
-\frac{ \beta (4 \pi)^k }{8 \pi \ell^2 } \Big[ m \ell^2 \\ 
+ d \Big( \sum_{i=0}^k \binom{k}{i} \frac{ (-1)^i N^{2i} \rho_+^{2k-2i+1} }{
2k-2i+1 } \Big) \Big]%
\end{array}
$ \\ 
&  &  &  &  \\ \hline
\end{tabular}%
\end{table}

\begin{table}[!hbp]
\caption{Summary of Entropies}
\label{summarytableS}\centering        
\begin{tabular}{|c|c|c|}
\hline
dim. & R-approach & C-approach \\ \hline
&  &  \\ 
4 & $\frac{ \beta ( \tau_c^3 + 3 n^2 \tau_c - m \ell^2 )}{2\ell^2} $ & $%
\frac{ \beta ( \rho_+^3 - 3 N^2 \rho_+ - m \ell^2 )}{2\ell^2} $ \\ 
&  &  \\ \hline
&  &  \\ 
6 & $%
\begin{array}{c}
\frac{2 \pi \beta }{3\ell^2} \Big( 3 \tau_c^5 + 10 n^2 \tau_c^3 \\ 
+ 15 n^4 \tau_c - 9 m \ell^2 \Big)%
\end{array}
$ & $%
\begin{array}{c}
\frac{2 \pi \beta }{3\ell^2} \Big( 3 \rho_+^5 - 10 N^2 \rho_+^3 \\ 
+ 15 N^4 \rho_+ - 9 m \ell^2 \Big)%
\end{array}
$ \\ 
&  &  \\ \hline
&  &  \\ 
8 & $%
\begin{array}{c}
\frac{ 8 \pi^2 \beta }{5\ell^2} \Big( 5 \tau_c^7 + 21 n^2 \tau_c^5 \\ 
+ 35 n^4 \tau_c^3 + 35 n^6 \tau_c \\ 
- 25 m \ell^2 \Big)%
\end{array}
$ & $%
\begin{array}{c}
-\frac{ 8 \pi^2 \beta }{5\ell^2} \Big( -5 \rho_+^7 + 21 N^2 \rho_+^5 \\ 
- 35 N^4 \rho_+^3 + 35 N^6 \rho_+ \\ 
+ 25 m \ell^2 \Big)%
\end{array}
$ \\ 
&  &  \\ \hline
&  &  \\ 
10 & $%
\begin{array}{c}
\frac{ 32 \pi^3 \beta }{35 \ell^2 } \Big( 35 \tau_c^9 + 180 n^2 \tau_c^7 \\ 
+ 378 n^4 \tau_c^5 + 420 n^6 \tau_c^3 \\ 
+ 315 n^8 \tau_c - 245 m \ell^2 \Big)%
\end{array}
$ & $%
\begin{array}{c}
\frac{ 32 \pi^3 \beta }{35 \ell^2 } \Big( 35 \rho_+^9 - 180 N^2 \rho_+^7 \\ 
+ 378 N^4 \rho_+^5 - 420 N^6 \rho_+^3 \\ 
+ 315 N^8 \rho_+ - 245 m \ell^2 \Big)%
\end{array}
$ \\ 
&  &  \\ \hline
&  &  \\ 
(d+1) & $%
\begin{array}{c}
\frac{(4\pi)^k \beta }{8 \pi \ell^2} \Big[ d \sum_{i=0}^k \binom{k}{i} \frac{
n^{2i} \tau_c^{2k - 2i + 1} }{2k - 2i + 1 } \\ 
- m \ell^2 (2k-1) \Big]%
\end{array}
$ & $%
\begin{array}{c}
\frac{(4\pi)^k \beta }{8 \pi \ell^2} \Big[ d \sum_{i=0}^k \binom{k}{i} \frac{
(-1)^i N^{2i} \rho_+^{2k - 2i + 1} }{2k - 2i + 1 } \\ 
- m \ell^2 (2k-1) \Big]%
\end{array}
$ \\ 
&  &  \\ \hline
\end{tabular}%
\end{table}

\end{document}